\documentclass[twocolumn,prb,showpacs,superscriptaddress,longbibliography]{revtex4-1}
\usepackage[utf8]{inputenc}
\usepackage{amssymb}
\usepackage{amsmath}
\usepackage{bbm}
\usepackage{graphicx}
\usepackage[colorlinks=true,citecolor=blue]{hyperref}
\usepackage[export]{adjustbox}

\newcommand{\dga}{D$\Gamma$A}

\renewcommand{\vec}[1]{\boldsymbol{#1}}
\newcommand{\ira}{\mathcal{I}}
\newcommand{\mas}{\mathcal{M}}
\newcommand{\pprop}{\mathcal{P}}
\newcommand{\pacc}{\mathcal{A}}
\newcommand{\accratio}{R}
\newcommand{\up}{\uparrow}
\newcommand{\down}{\downarrow}
\newcommand{\sgn}{\mathrm{sgn}}

\newcommand{\eg}{e.g.\ }
\newcommand{\ie}{i.e.\ }
\newcommand{\etal}{\textit{et al.}\ }

\newcommand{\etc}{etc.\ }

\newcommand\Let{\mathrel{\mathop:\!\!=}}

\newcommand{\cpxi}{{\ensuremath{i}}}

\newcommand{\inu}{{\ensuremath{\cpxi\nu}}}
\newcommand{\kv}{\ensuremath{\mathbf{k}}}

\newcommand{\av}[1]{\ensuremath{\left\langle #1 \right\rangle}}
\DeclareMathOperator{\Tr}{Tr}
\let\Re\relax
\let\Im\relax
\DeclareMathOperator{\Re}{Re}
\DeclareMathOperator{\Im}{Im}

\begin{document}

\title{Diagrammatic Monte Carlo approach for diagrammatic extensions of dynamical mean-field theory -- convergence analysis of the dual fermion technique}

\author{Jan Gukelberger}
\email[Corresponding author: ]{j.gukelberger@usherbrooke.ca} 
\affiliation{D\'epartement de Physique and Institut quantique, Universit\'e de Sherbrooke, Sherbrooke, Qu\'ebec, J1K 2R1, Canada}
\author{Evgeny Kozik}
\affiliation{Physics Department, King's College London, Strand, London WC2R 2LS, United Kingdom}
\author{Hartmut Hafermann}
\affiliation{Mathematical and Algorithmic Sciences Lab, France Research Center, Huawei Technologies France SASU, 92100 Boulogne-Billancourt, France} 

\date{\today}

\hyphenation{}

\begin{abstract}
The dual-fermion approach provides a formally exact prescription for calculating properties of a correlated electron system in terms of a diagrammatic expansion around dynamical mean-field theory (DMFT). 
Most practical implementations, however, neglect higher-order interaction vertices beyond two-particle scattering in the dual effective action and further truncate the diagrammatic expansion in the two-particle scattering vertex to a leading-order or ladder-type approximation. In this work we compute the dual-fermion expansion for the two-dimensional Hubbard model including all diagram topologies with two-particle interactions to high orders by means of a stochastic diagrammatic Monte Carlo algorithm. We benchmark the obtained self-energy against numerically exact Diagrammatic Determinant Monte Carlo simulations to systematically assess convergence of the dual-fermion series and the validity of these approximations. We observe that, from high temperatures down to the vicinity of the DMFT Néel transition, the dual-fermion series converges very quickly to the exact solution in the whole range of Hubbard interactions considered  ($4 \leq U/t \leq 12$), implying that contributions from higher-order vertices are small. As the temperature is lowered further, we observe slower series convergence, convergence to incorrect solutions, and ultimately divergence. This happens in a regime where magnetic correlations become significant. We find however that the self-consistent particle-hole ladder approximation yields reasonable and often even highly accurate results in this regime.
\end{abstract}


\maketitle

\section{Introduction}

Strongly correlated fermion systems pose a formidable challenge in contemporary condensed matter physics. Due to the complexity of the problem, these systems are primarily studied via numerical methods. Dynamical mean-field theory~\cite{Georges1996} (DMFT) has provided important insights, notably into the Mott metal-insulator transition.
However, as a mean-field theory it neglects spatial correlations altogether and sometimes fails to describe the physics even qualitatively, in particular in low-dimensional systems.
Cluster extensions of DMFT improve upon DMFT by including all the contributions to the self-energy that are local to the cluster.~\cite{maier2005qct} The range of the spatial correlations is determined by the cluster size as a control parameter. The solution becomes exact in the limit of infinite cluster size. Because of the exponentially growing complexity however, convergence with respect to the cluster size is seldom reached in practice.

Diagrammatic extensions of DMFT are complementary to cluster extensions. 
Unlike clusters, correlations can be included in the self-energy up to all length scales within certain approximations.
A common feature underlying these methods is that in addition to the local self-energy, they utilize higher-order moments of the impurity, more precisely vertex functions, to construct diagrammatic approximations to the momentum dependent self-energy.
Examples comprise the dynamical vertex approximation (\dga),~\cite{Toschi2007} the dual fermion (DF),~\cite{Rubtsov2008} dual boson (DB),~\cite{Rubtsov2012} the one-particle irreducible approach (1PI),~\cite{Rohringer2013} and the triply irreducible local expansion (TRILEX).~\cite{Ayral2015}

For a given approach, different levels of approximation to the self-energy can be constructed.
Almost all calculations, including those performed here, neglect vertex functions beyond the two-particle level because the cost for computing functions of six or more frequency arguments quickly becomes prohibitive.
Within this approximation, the parquet formalism allows to determine both the self-energy and vertex function self-consistently and incorporates important contributions from all scattering channels and their interaction. Its feasibility has been demonstrated for both \dga~\cite{Valli2015} and DF,~\cite{Yang2011a} but the substantial computational effort limits applications to systems consisting of a small number of lattice sites.
Alternatively, diagrams are chosen based on physical considerations. An important approximation is the (particle-hole) ladder approximation.
In the two-dimensional Hubbard model, this approximation captures the effects of collective spin and charge excitations on the self-energy and has been shown to describe pseudogap physics.~\cite{Hafermann2009} In certain regimes, where these fluctuations are dominant, the approximation can even yield quantitatively accurate results. Examples can be found in the literature for self-energies~\cite{Leblanc2015} and critical exponents,~\cite{Hirschmeier2015} and in the results section of this paper.
The restriction to a subset of diagrams introduces a bias and potentially neglects relevant effects. The applicability of such approximations is expected to be limited to certain parameter regimes. Furthermore, error bars of the solution are unknown.

In this work we propose a method that augments diagrammatic extensions of DMFT with a controlled way of summing the diagrammatic corrections to the DMFT self-energy by means of a diagrammatic Monte Carlo (DiagMC) algorithm.
In principle, the method samples diagrams of all allowed topologies and does not require any momentum discretization.
While our method can be applied to any kind of diagrammatic extension such as \dga, DF, DB, 1PI, or TRILEX and to different models, we focus here on the dual fermion approach to the Hubbard model on a square lattice as the first application (DiagMC@DF). In our implementation of DiagMC@DF we take into account all topologies of the DF diagrams involving only the two-particle scattering vertex of the dual effective action. As mentioned above, handling higher-order vertices of the theory is currently infeasible and is left for future work. The contribution of neglected diagrams is expected to be small in certain regimes, but can not be accessed \textit{a priori}. Therefore, benchmarking of the obtained results against numerically exact Diagrammatic Determinant Monte Carlo (DDMC) calculations performed here is crucial for assessing applicability of the new method. Correspondingly, the error bars claimed for our DiagMC@DF results express the controlled statistical and systematic uncertainty of the dual theory truncated at two-particle scattering. The combined contribution due to three-particle and higher-order vertices can be evaluated as deviation from the benchmark.     

The structure of the paper is as follows: 
In Sec.~\ref{sec:overview} we give an overview of our results and discuss them in relation to convergence properties of diagrammatic series. Section~\ref{sec:method} shortly reviews the dual-fermion approach for the Hubbard model before giving a detailed description of the DiagMC@DF algorithm. We also describe the essentials of the DDMC method used for benchmarks.
Section~\ref{sec:results} contains the results of our benchmarks exploring the reliability of the DiagMC@DF method and of different approximation schemes.
We wrap up with a discussion of our findings and future directions in Sec.~\ref{sec:discussion} and a brief summary in Sec.~\ref{sec:conclusions}.

\section{Convergence of diagrammatic series and overview of results}
\label{sec:overview}

In diagrammatic theories where the answer is formally represented by a sum of an infinite series of terms, understanding convergence properties of the series is key to claiming a controlled result. Using DiagMC techniques one can evaluate generic sign-alternating diagrammatic series for fermionic models, in which the number of diagrams grows factorially with diagram order, up to order $N_* \sim 10$. At orders greater than $N_*$ the statistical noise due to the sign alternation explodes and obtaining meaningful results is not possible. Thus, for the theory to produce a reliable result the series must exhibit manifest convergence---a plateau of the result within statistical errors as a function of diagram order or at least a fast decay of higher-order corrections---at orders below $N_*$ or be practically resummable, i.e.\ amenable to analytic continuation beyond convergence radius given the data up to order $N_*$. 

Divergence of a diagrammatic series naturally stems from physical non-analyticity of observables as is for example the case at second-order phase transitions -- the sum of all terms in the series is bound to diverge at the transition point. If, in contrast, the sum is restricted to specific diagram topologies only, such a series can diverge in the normal phase well before the actual phase transition. A textbook example is the sum of all ladder diagrams in a fluctuating channel, which, being the set of fastest growing diagrams in the weak-coupling limit, typically diverges above the actual critical point. In the case of the DF theory for the half-filled 2D Hubbard model, such a ladder sum in the particle-hole channel built on the \emph{bare} dual propagators and two-particle vertex diverges at the DMFT N\'eel temperature $T_N^\mathrm{DMFT}$,~\cite{Otsuki2014} whereas the actual antiferromagnetic order sets in at $T_N=0$ (the ladder constructed from renormalized propagators does not exhibit this divergence). Nonetheless, this does not imply \textit{a priori} that the bare series of all diagram topologies sampled by DiagMC@DF is meaningless below $T_N^\mathrm{DMFT}$. On the contrary, already in moderately correlated regimes, divergence of ladder diagrams is typically compensated by growing terms of other topologies so that the whole sum is insensitive to the spurious divergence of the ladders.~\cite{Gukelberger2015}

However, divergence of diagrammatic series can also be unrelated to physics. For instance, in models with Hubbard interaction, an involved analytic structure of the self-energy in the complex plane of the interaction parameter $U$ makes the diagrammatic series in terms of $U$ and the non-interacting Green's function diverge in the vicinity of half-filling. For the doped Hubbard model with hopping $t$ this leads to failure of DiagMC based on the bare series at temperatures $T \lesssim 0.5 t$ and $U \gtrsim 4t$.~\cite{Wu2016} A very similar divergence is exhibited by the Hubbard atom---the simplest model with Hubbard interaction and $t=0$, which features no non-trivial physics, let alone a phase transition. An analysis of the self-energy for the Hubbard atom reveals that the divergence is due to multiple poles in the complex $U$-plane, which have no clear physical origin.~\cite{Wu2016} There is very little intuition on where to expect such divergences and a systematic study of the series is the only practical approach to controlling the answer. 

In this paper we perform such a study of the series evaluated by DiagMC@DF (summarised in Fig.~\ref{fig:phasediag} for the half-filled model). We find that convergence properties of the dual bare series are considerably enhanced compared to that of the original Hubbard model. In particular at temperatures $T \sim 0.5 t$ the DF series converges within order $6$ up to the highest value of interaction attempted, $U=12t$. By benchmarking against DDMC we verify that DiagMC@DF reproduces the exact self-energy in this regime, implying that the contribution due to neglected high-order vertices is indeed small. Here and below by convergence to the exact result we mean that high-order results for the self-energy and the benchmark agree within 1\% of the absolute value.

Considerably worse than definite divergence of a series is the case when it converges to a wrong answer. Such a situation was observed recently in models with Hubbard interaction for skeleton series built on the full interacting Green's function and was related to multivaluedness of the corresponding Luttinger-Ward functional.~\cite{kozik2015nlf} Just like divergence of the bare series, the misleading convergence of the skeleton series happens in the Hubbard model around half filling, $T \lesssim 0.5 t$, and $U \gtrsim 4t$ and is not obviously linked to any physics -- the trivial Hubbard atom and all other models with the Hubbard interaction considered in Ref.~\onlinecite{kozik2015nlf} behave in the same way, consistently with universality of the Luttinger-Ward functional. As a result, none of the current DiagMC approaches based on standard Feynman diagrammatics, neither bare nor skeleton, can controllably access the regime of strong coupling of $U \gtrsim 4t$ in the vicinity of half-filling at low temperatures.~\footnote{Some progress in reaching strong correlations by DiagMC has been made in Ref.~\onlinecite{Wu2016}, where the convergence radius of the bare series was extended by introducing an auxiliary field.} 

We find that DiagMC@DF, in contrast, can reliably reach $U=12 t$ down to $T \sim 0.5 t$. However, as the temperature is lowered further down to the DMFT Néel transition the convergence of the evaluated DF series becomes worse and at $U \gtrsim 8t$ it clearly converges to a wrong answer. Although we cannot rigorously rule out the misleading convergence scenario of Ref.~\onlinecite{kozik2015nlf}, it is quite unlikely since the evaluated series is constructed in terms of the non-interacting dual Green's function, which has no direct connection to the Luttinger-Ward functional. Moreover, DiagMC@DF accurately reproduces the exact result in the regime ($T \sim 0.5 t$, $U \gtrsim 4t$) where the standard skeleton expansion exhibits misleading convergence to the unphysical branch of the Luttinger-Ward functional. In the case of DiagMC@DF, where only the part of the dual action truncated at two-particle scattering is solved in a controlled way, there is a more natural explanation for the discrepancy -- the omitted series involving higher-order scattering vertices can naturally give a substantial contribution to the answer. Our results, therefore, allow to assess the validity of the standard approximation involved in the conventional DF approach, namely the truncation of the dual fermion coupling to two-particle interactions. 

As the temperature reaches the regime of extended (to about $4$ lattice sites) magnetic correlations, the DF series is found to diverge. It seems plausible, at least in principle, that the divergence could be a result of the truncation of the effective action at two-particle scattering and the omitted series of diagrams could compensate the divergence. A separate study is necessary to answer this question.   

We further asses the accuracy of the standard DF particle-hole ladder approximation. We find that the self-consistent ladder, which is known to converge far below $T_N^\mathrm{DMFT}$ due to self-consistent dressing of the dual particle-hole bubble,~\cite{Otsuki2014}  produces meaningful results in certain regimes where the bare DF series is divergent. This suggests that a self-consistent DiagMC@DF approach built on dressed dual Green's functions could potentially extend the applicability range of the technique, but this possibility requires a separate investigation. Our results also shed light on the role of the mapping from the dual to the physical self-energy.

Finally, it should be noted that our method is fundamentally different from the earlier proposal of combining DMFT with DiagMC.~\cite{Pollet2011} In that approach DiagMC sums non-local skeleton diagrams for the original model where at least one Green's function connects different lattice sites. The sum of all the purely local graphs is obtained from the solution of the DMFT impurity problem by non-perturbative means, such as a Monte Carlo impurity solver. However, being based on the skeleton diagrammatic technique for the original lattice fermions, the approach of Ref.~\onlinecite{Pollet2011} fails for the Hubbard model in correlated regimes because of the convergence of the skeleton series to a wrong result as discussed above.

\section{Model and method}\label{sec:method}

To set the stage for the description of the DiagMC@DF method, we first recapitulate the DF approach. Readers familiar with this method may skip to Sec.~\ref{sec:DiagMC@DF}, where we explain the DiagMC algorithm adapted to sample DF diagrams.

To avoid confusion, we note that our calculations involve two kinds of Monte Carlo sampling: we employ a hybridization expansion continuous-time quantum Monte Carlo (CTHYB) solver~\cite{Werner06} for the solution of the underlying impurity model and diagrammatic Monte Carlo sampling of the dual-fermion diagrams.
In the following we only discuss the latter as it defines our method.
The impurity solver can be replaced by any other method such as another flavor of CTQMC or exact diagonalization. We expect our conclusions to be independent of the choice of impurity solver as long as it is numerically exact.

For definiteness we consider the single-band Hubbard model on the two-dimensional square lattice both at and away from half filling. Its Hamiltonian is given by~\footnote{Following standard notation, $c_{i\sigma}$ and $c_{i\sigma}^{\dagger}$ denote annihilation and creation operators for fermions at site $i$ with spin projection \unexpanded{$\sigma=\uparrow,\downarrow$}, $\kv$ and $i$ label momenta and sites, and $U$ and $\mu$ denote the Hubbard interaction and chemical potential, respectively.}
\begin{equation}
\label{eq:H}
H=  \sum_{\kv\sigma} \epsilon_{\kv}c^\dagger_{\kv\sigma}c_{\kv\sigma}+\sum_i [Un_{i\uparrow}n_{i\downarrow}-\mu (n_{i\uparrow}+n_{i\downarrow})]
\end{equation}
with bare dispersion $\epsilon_{\kv}=-2t(\cos k_{x}+\cos k_{y})$. The nearest-neighbor hopping $t = 1$ sets the unit of energy for the remainder of this paper.

\subsection{Dual fermions} \label{sec:df}

We briefly sketch the derivation of the DF method, concentrating on the arguments relevant for the subsequent discussion of the DiagMC@DF method. Further details on the DF method can be found in the literature.~\cite{Rubtsov2008,Hafermann2009a}

DF is essentially a perturbation expansion around an impurity model which serves as a solvable reference system. The underlying idea is to treat the strong local correlations on the level of the single-site impurity model and to embrace the presumably weaker coupling between the sites perturbatively. A diagrammatic extension of DMFT is obtained by setting the hybridization function to its DMFT value.

In the Grassmann path integral formalism, the impurity model is described by the action
\begin{align*}
S_{\text{imp}}[c^{*},c]=-\!\!\sum_{\nu\sigma} c^{*}_{\nu\sigma}[\inu+\mu-\Delta_{\nu\sigma}]c_{\nu\sigma}\!+ U\sum_{\omega}n_{\omega\uparrow}n_{-\omega\downarrow},
\end{align*}
where $\nu$ is the Matsubara frequency and $\Delta_{\nu\sigma}$ is the hybridization function, which is arbitrary at this point. The associated impurity Green's function is defined in the standard way, $g_{\nu\sigma} \Let -\av{c_{\nu\sigma} c^{*}_{\nu\sigma}}$, where $\av{\ldots}$ is the average with respect to the action $S_{\text{imp}}[c^{*},c]$.
By adding and subtracting the hybridization function, the lattice action corresponding to the Hamiltonian~\eqref{eq:H} can be written in the form
\begin{align*}
S_{\text{latt}}[c^{*},c]=\sum_{i}S_{\text{imp}}[c_{i}^{*},c_{i}] &- \sum_{\kv\nu\sigma}c^{*}_{\kv\nu\sigma}(\Delta_{\nu\sigma}-\varepsilon_{\kv})c_{\kv\nu\sigma}.
\end{align*}
The decoupling of the second term in this equation is achieved through an exact integral transformation, which leads to the partition function
\begin{align}
\label{partitionfunction}
Z =  D_{f} &\int \mathcal{D}[f^{*},f] e^{-\sum_{\kv\nu\sigma}f^{*}_{\kv\nu\sigma}g_{\nu\sigma}^{-1}(\Delta_{\nu\sigma}-\varepsilon_{\kv})^{-1}g_{\nu\sigma}^{-1}f_{\kv\nu\sigma}}\times\notag\\
&\int  \mathcal{D}[c^{*},c]  e^{-\sum_{i}\{ S_{\text{imp}}[c^{*}_{i},c_{i}] + S_{\text{cf}}[c^{*}_{i},c_{i};f^{*}_{i},f_{i}]\}}.
\end{align}
Here $D_{f}$ is a determinant which is irrelevant for the calculation of expectation values. The term $S_{\text{cf}}$ is the local coupling between dual and physical fermions:
\begin{align*}
S_{\text{cf}}[c^{*},c;f^{*},f] =& \sum_{\nu\sigma} \left(f^{*}_{\nu\sigma}g_{\nu\sigma}^{-1}c_{\nu\sigma}+c^{*}_{\nu\sigma}g_{\nu\sigma}^{-1}f_{\nu\sigma}\right).
\end{align*}
The physical fermionic degrees of freedom in the second line of \eqref{partitionfunction} can be integrated out for each site separately after expanding in $S_{\text{cf}}$. Because of the $S_{\text{imp}}$ in the exponential, the integral over $c$, $c^{*}$ corresponds to taking the impurity average. The result can be expressed in the following form:
\begin{align}
\label{Vdef}
\ln \av{e^{-S_{\text{cf}}[c^{*},c;f^{*},f]}}=&-\sum_{\nu\sigma}f_{\nu\sigma}^{*}g^{-1}_{\nu\sigma}f_{\nu\sigma}- \tilde{V}[f^{*},f].
\end{align}
The left hand side generates the \emph{connected} correlation functions of the impurity model coupled to dual variables. To leading order, the resulting dual interaction is given by
\begin{align}
\label{V}
&\tilde{V}[f^{*},f] = - \frac{1}{4}\!\!\!\! \sum_{\nu\nu'\omega\,\sigma_{i}}\!\!\!\!\gamma^{\sigma_{1}\sigma_{2}\sigma_{3}\sigma_{4}}_{\nu\nu'\omega}f_{\nu\sigma_{1}}^{*}f_{\nu+\omega,\sigma_{2}}f^{*}_{\nu'+\omega,\sigma_{3}}f_{\nu'\sigma_{4}}\!\!+\!\ldots,
\end{align}
where $\gamma$ is the reducible two-particle vertex of the impurity model.
The higher-order terms contain the three-particle (six-leg) and higher-order vertices of the impurity.

Combining Eqs.~\eqref{partitionfunction} and \eqref{Vdef}, we obtain the action in dual variables:
\begin{align}
\tilde{S}[f^{*},f] = &- \sum_{\kv\nu\sigma}f^{*}_{\kv\nu\sigma} [\tilde{G}^{(0)}_{\kv\nu\sigma}]^{-1} f_{\kv\nu\sigma} + \tilde{V}[f^{*},f].\label{dualaction}
\end{align}
The bare dual Green's function can be identified by looking at the bilinear terms in the same two equations:
\begin{align}
\label{gdual}
\tilde{G}^{(0)}_{\kv\nu\sigma} =&\left[g_{\nu\sigma}^{-1} + (\Delta_{\nu\sigma}-\varepsilon_{\kv})\right]^{-1} - g_{\nu\sigma}.
\end{align}
All steps so far are free from approximations. Exact relations between the physical and dual quantities can therefore be established. In particular, the physical self-energy is given in terms of the dual self-energy $\tilde{\Sigma}$ by
\begin{align} \label{eq:sigma-from-sigmad}
\Sigma_{\kv\nu\sigma} = \Sigma^{\text{imp}}_{\nu\sigma} + \frac{\tilde{\Sigma}_{\kv\nu\sigma}}{1+\tilde{\Sigma}_{\kv\nu\sigma}g_{\nu\sigma}}.
\end{align}

Up to now the hybridization function is arbitrary. Its value is generally fixed through the self-consistency condition $\sum_{\kv}\tilde{G}_{\kv\nu\sigma}=0$. We refer to these calculations as self-consistent. 
In one-shot (\ie non-self-consistent~\footnote{By self-consistent we refer to the self-consistent determination of the hybridization function as opposed to the self-consistent renormalization of dual Green's functions in the diagrams. In practical dual fermion approximations that we quote here, diagrams are typically summed with self-consistently renormalized propagators.}) dual-fermion calculations this condition reduces to $\sum_{\kv}\tilde{G}^{(0)}_{\kv\nu\sigma}=0$. It has been shown~\cite{Rubtsov2008} that the solution to this latter equation is $\Delta_{\nu\sigma}=\Delta^{\text{DMFT}}_{\nu\sigma}$. In that case, Eq.~\eqref{gdual} becomes equal to $G^{\text{DMFT}}_{\kv\nu\sigma}-g_{\nu\sigma}$, which establishes the connection to DMFT. 
The dual-fermion expansion can therefore be understood as an expansion around the DMFT solution, in the sense that a vanishing dual self-energy leads to the DMFT solution $\Sigma_{\kv \nu \sigma} = \Sigma_{\nu \sigma}^{\text{DMFT}}$. Higher orders of dual-fermion perturbation theory yield non-local corrections to the self-energy.

\subsubsection{Convergence properties}
\label{sec:convergence}

Equations~\eqref{V}-\eqref{gdual} define an interacting fermionic lattice model which is usually solved within perturbation theory. Despite the formal equivalence to the original model~\eqref{eq:H}, one can expect different convergence behavior for expansions in the original or the dual interaction because, in the dual system, the presumably strong local correlations are already taken into account on the level of the impurity model.
In the weak-coupling limit the DF series converges fast since $\gamma\sim U$.  In the strong coupling limit, the dual Green's function plays the role of  a small parameter. The argument is that for strong interaction, the system is close to the atomic limit, so that $\Delta$ and $\epsilon_{\kv}$ and therefore $\tilde{G}^{(0)}$ are small. Fast convergence in both limits has been observed numerically, based on the leading eigenvalue of the Bethe-Salpeter equation in the particle-hole channel.~\cite{Hafermann2009} A smaller eigenvalue implies faster convergence of the ladder diagram series.
Here we use the DiagMC@DF approach to assess convergence properties of the whole series in $\gamma$, including all possible diagram topologies.

\subsubsection{Common approximations}

While all steps in the derivation of the dual model are exact, two approximations are usually employed in practical calculations based on the DF formalism: i) the DF interaction $\tilde{V}$ is truncated after the leading-order term, restricting it to the two-particle vertex $\gamma$, and ii) in summation of the diagrammatic series for the dual self-energy in terms of $\gamma$ only a subset of diagrams is taken into account.

The first approximation is made because the computational effort of computing and handling the $n$-particle interaction terms grows catastrophically (exponentially with the base given by the frequency mesh size) with $n$. On physical grounds, one may expect that three-particle scattering processes are less relevant than two-particle processes. In Ref.~\onlinecite{Hafermann2009}, the corrections due to diagrams containing the three-particle vertex were found to be much smaller than those due to ladder diagrams.
However no rigorous argument exists. Here, by evaluating the whole diagrammatic series in terms of the two-particle vertex $\gamma$ with DiagMC@DF we can assess the accuracy of this approximation against numerically exact DDMC benchmarks.

The dual self-energy is most commonly evaluated in second-order or ladder diagram approximation. The former describes dynamical short-range antiferromagnetic correlations,~\cite{Brener2008} while the latter additionally accounts for pseudogap physics.~\cite{Hafermann2009} The often made restriction to particle-hole ladder diagrams is based on physical considerations: in the repulsive Hubbard model, these describe the dominant processes  at half filling and intermediate temperatures. The DiagMC@DF method allows us to remove the restriction of diagram topologies and verify the validity of such approximations in different parameter regimes.

\subsection{Diagrammatic Monte Carlo for dual fermions}
\label{sec:DiagMC@DF}

Diagrammatic Monte Carlo describes a general approach of computing a diagrammatic expansion via stochastic sampling with a Markov chain Monte Carlo algorithm.~\cite{prokofev2008fpp} Instead of a deterministic sum, the sum over all diagram topologies as well as the sums or integrals over the internal variables of each diagram are performed stochastically. Diagrams are sampled up to a fixed diagram order cutoff $N_{*}$.
A DiagMC configuration consists of a given diagram topology with all the internal variables (such as momenta, times, and spin indices) fixed to specific values, while a set of ergodic updates enables transitions between different topologies and/or changes of the internal variables with probabilities determined by the value of the diagram of each configuration. During the last ten years, the approach has found diverse applications such as polaron problems,~\cite{prokofev2007bdm} the resonant Fermi gas,~\cite{VanHoucke2012} correlated lattice fermions,~\cite{VanHoucke2010, deng2014ebr} and frustrated spin systems.~\cite{kulagin2013bdm2} In the case of the Hubbard model, the method has been shown to give reliable results in the regime of moderate interactions $U \lesssim W/2$, where $W$ denotes the bandwidth of the system,~\cite{kozik2010, Leblanc2015, gukelberger2015diss} but at larger interactions and relevant temperatures the diagrammatic series typically diverges so quickly that no physical results can be extracted.
This finding may not be surprising, since the method evaluates a perturbative expansion around the model's non-interacting limit (albeit to high order), which is in general not a good reference for the strongly interacting system.
In contrast, in this work we use the DiagMC approach to evaluate the dual-fermion expansion. As discussed in Sec.~\ref{sec:convergence}, it may be expected to converge quickly whenever the impurity model constitutes a good reference for the lattice model's exact solution, in particular in both the weak- and strong-coupling limits.
Most of the sampling algorithm is a straightforward generalization of the conventional DiagMC method for the Hubbard model, which has been described before in detail in Refs.~\onlinecite{VanHoucke2010, gukelberger2015diss}. In the following we give a self-contained description of the algorithm's essential steps because readers from the DF community may not be familiar with DiagMC, while some peculiarities of DF diagrammatics that need to be accounted for in the sampling process may not be obvious to DiagMC practitioners.

\subsubsection{Configurations}

In the present work, we focus on computing the dual self-energy $\tilde{\Sigma}_{\kv \nu \sigma}$ by sampling the corresponding DF diagrams with diagrammatic Monte Carlo. The result determines the lattice self-energy via Eq.~\ref{eq:sigma-from-sigmad} and hence all single-particle properties. The DF diagrams that define the set of valid configurations consist of one or more four-point vertices $\gamma_{\nu\nu'\omega}^{\sigma_1 \sigma_2 \sigma_3 \sigma_4}$ connected by dual propagator lines $\tilde{G}^{(0)}_{\kv\nu\sigma}$. In any given configuration, all spin, momentum, and frequency indices are fixed to specific values, such that the complex weight of the configuration is directly given by the product of the vertices and propagators at these indices. The values of these diagrammatic building blocks are usually determined by an initial DMFT calculation, which yields the impurity vertex $\gamma^{\text{imp}}$, hybridization function $\Delta$, and impurity Green's function $g$ and hence fixes the dual propagator \eqref{gdual}. 

We note that, unlike typical second-order and ladder DF calculations, we do not use a diagrammatic theory with symmetrized vertices, i.e.\ in our case all propagators connected to a given interaction vertex are considered distinct for the purpose of enumerating diagram topologies. A consequence is that the value of the interaction vertex $\gamma$ in our diagrams differs from the impurity vertex $\gamma^{\text{imp}}$ by a constant factor to avoid double counting: $\gamma = \frac{1}{4} \gamma^{\text{imp}}$. For a more detailed explanation of this subtlety the reader is referred to appendix \ref{sec:symmetrization}.

For numerical convenience, we choose to work in momentum and imaginary-frequency space because here the vertices are easily representable by a table for the different frequency and spin configurations. Working in imaginary-time space would have the advantage of purely real diagram values but require interpolation of the vertex in imaginary time and careful treatment of nonanalyticities in the vertex's time-dependence.

The sampled self-energy diagrams formally have two open ends (truncated in- and out-going lines), but it is convenient~\cite{VanHoucke2010} for handling the configurations to use an internal representation where the open ends are connected with a dummy propagator whose weight is one, so that the diagram has no open ends. Since this dummy line carries the momentum, frequency and spin index of the self-energy diagram, which need to be queried at each measurement, one refers to it as the ``measuring propagator''.

Since momentum and frequency conservation would prevent any local update from changing the momentum- and frequency indices of a line or vertex, we need to enlarge the configuration space to include unphysical diagrams where 4-momentum conservation is violated in two locations. In the spirit of the seminal worm algorithm,~\cite{prokofev1998worm} following Ref.~\onlinecite{VanHoucke2010}, we call these two locations ``worms'', denote them by the symbols $\ira$ (``Ira'') and $\mas$ (``Masha''), and say that an excess 4-momentum $\delta$ flows from $\mas$ to $\ira$.

Additionally, we include an artificial normalization diagram in the configuration space that has the form of a Hartree diagram where the interaction vertex is replaced by a dummy vertex with unit weight. The only nontrivial element of this diagram is hence a single dual propagator, whose integral $\int (dk) |\tilde{G}^{(0)}_k|$ is straightforwardly evaluated numerically. Together with the number of times $M_{\mathrm{norm}}$ this diagram is measured, the normalization factor relating measurements to diagram value is then known as $\mathcal{N} = \int (dk) |\tilde{G}^{(0)}_k| / M_{\mathrm{norm}}$.

Since the value of a configuration is a complex number, we need to store its complex sign (or phase) along with the configuration. Accepted Monte Carlo updates generally update the sign by the phase factor corresponding to the move: $s \leftarrow s \cdot \sgn(w_{c'}/w_c)$ with the complex sign function $\sgn (z) \equiv z / |z|$ and $w_c$ the complex weight of configuration $c$, \ie the product of all vertices and propagators in $c$. In order to avoid the accumulation of rounding errors, the sign is reset whenever the normalization diagram is encountered; in this case, the diagram's sign $\sgn(-\tilde{\Sigma}^{(0)}_{\kv \nu \sigma})$ is determined by the value of the only non-trivial propagator and a factor of -1 to account for the presence of one fermion loop.

\subsubsection{Updates} \label{sec:updates}

In the following we describe a sufficient set of Monte Carlo updates for sampling the diagrammatic space of dual self-energy diagrams ergodically. For every Monte Carlo step, one of these update types is randomly proposed with equal probabilities and accepted or rejected according to the Metropolis scheme. When stating proposal probabilities for specific moves, we do not include the implicit factor $1/N_{\mathrm{updates}}$ because it will drop out whenever an acceptance ratio is calculated. We use the symbols $\pprop$ for proposal and $\pacc$ for (complex) acceptance weights. The meaning of a complex $\pacc$ is that the update is accepted with probability $|\pacc|$ and in case of acceptance the sign of the new configuration will differ from the old one by a factor $\sgn(\pacc)$.

\begin{figure}[htp!]
    \centering
    \includegraphics[width=\columnwidth]{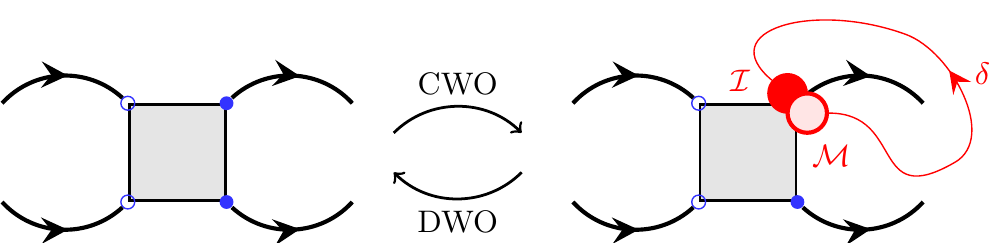}
    \caption{The CWO update creates a pair of worms with excess 4-momentum $\delta$ on a random vertex corner. The converse DWO move removes a pair of worms in the same location.}
    \label{fig:cwo}
\end{figure}

\paragraph{Create/Delete Worms (CWO/DWO)}
The simplest pair of updates creates transitions between the physical and worm sectors. If the diagram is already in the worm sector, the CWO update is trivially rejected, otherwise it creates a pair of worms on a random corner of a random vertex. For simplicity, the excess momentum $\vec{\delta}$ is drawn from a uniform distribution over the Brillouin zone $P_{\vec{\delta}} = U([-\pi,\pi)^d)$ and the excess frequency $\Delta$ from a discrete set of bosonic frequencies $-\omega_{N_\Delta}, \dots, \omega_{N_\Delta}$ with equal probabilities $P_\Delta = 1/(2N_{\Delta}+1)$. In practice a window width $\omega_{N_\Delta} \sim 10 \pi T$ is a good compromise between high acceptance probabilities and efficient changes of frequency indices. The proposal probability for a specific CWO move is hence $\pprop_{CWO} = P_{\vec{\delta}} P_\Delta / 4 n$, where $n$ denotes the number of vertices in the current diagram.

The converse DWO update is even simpler: If the worms $\ira$ and $\mas$ exist and are located on the same corner of a vertex, they are deleted and the diagram becomes physical again. Otherwise, the update is trivially rejected. We are therefore left with the trivial proposal probability $\pprop_{DWO}=1$.

The postulate of detailed balance between these updates implies the acceptance ratio
\begin{align}
\accratio_{CWO} &\equiv \frac{\pacc_{CWO}}{\pacc_{DWO}} 
= \frac{\pprop_{DWO}}{\pprop_{CWO}} C_W(n, \delta) \nonumber\\&
= \frac{4 n C_W(n, \delta)}{P_{\vec{\delta}} P_\Delta} .
\end{align}
Here we have made use of the fact that the unphysical worm sector can be associated with an arbitrary weight factor $C_W$, which can be tuned to improve sampling efficiency. In particular, the choice $C_W(n, \delta) = P_{\vec{\delta}} P_\Delta / 4 n$ ensures that \emph{all} allowed updates are accepted.

\begin{figure}[htp!]
    \centering
    \includegraphics[width=.9\columnwidth]{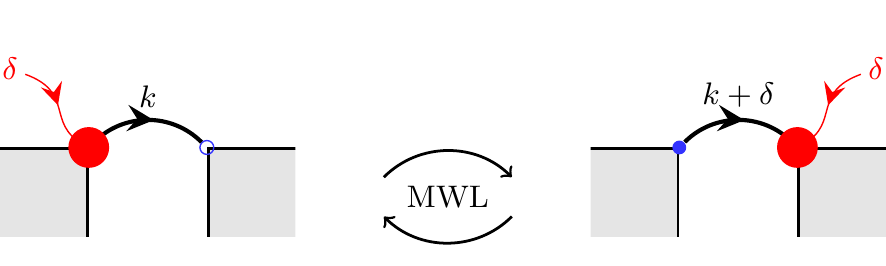}
    \includegraphics[width=\columnwidth]{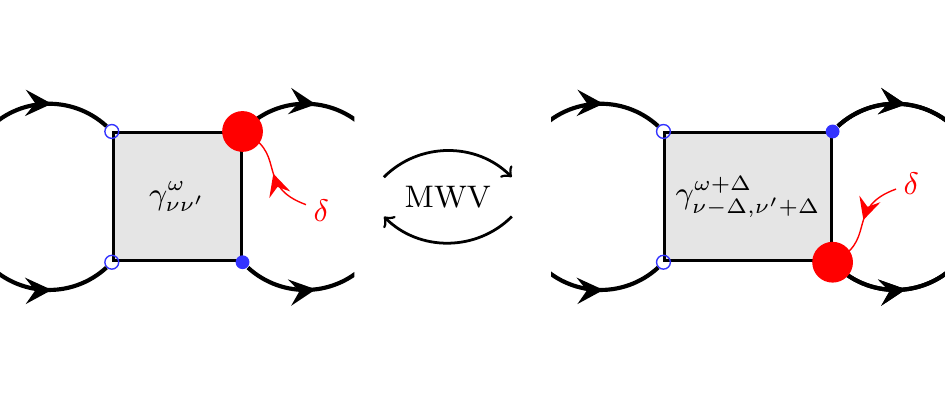}
    \includegraphics[width=\columnwidth]{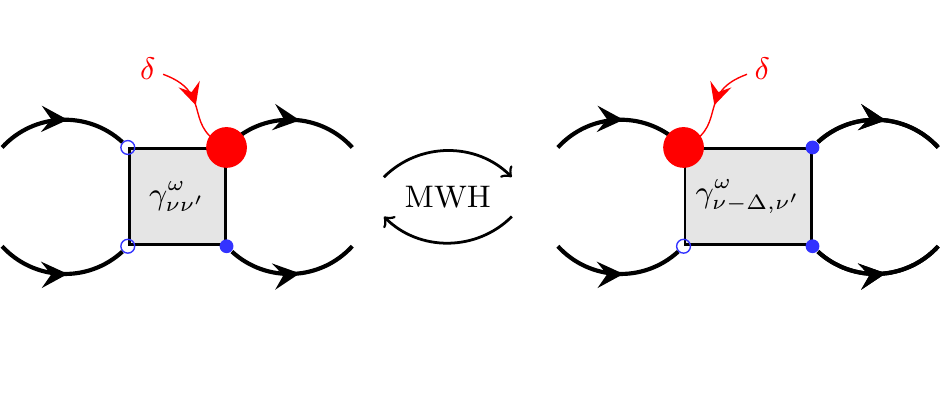}
    \caption{The MWL/MWV/MWH updates move a worm to a neighboring location and update momentum and frequency indices of the touched propagator or vertex accordingly. The pictures show the moving of $\ira$. Moving $\mas$ has the same effect except that the sign of the 4-momentum $\delta=(\Delta,\vec{\delta})$ is flipped.}
    \label{fig:mw}
\end{figure}

\paragraph{Move Worm along Line/Vertically/Horizontally (MWL/MWV/MWH)}
Once a pair of worms has been created, the diagram's 4-momentum indices can be changed by moving either of the worms to a neighboring vertex corner. We allow for three directions, namely along the propagator line that starts or ends in the current position (MWL), or to the neighboring corner of the current vertex in vertical direction (MWV), or to the horizontally neighboring corner (MWH). Each of these moves chooses one of the worms $\ira$ or $\mas$ with proposal probability 1/2 and is balanced with itself. The acceptance ratios are therefore directly given by the ratios between the new and old values of the element (propagator or vertex) whose 4-momentum indices are changed.

Implementation of these updates requires some care regarding a consistent treatment of the frequency indices of vertices with one or more adjacent worms. The simplest algorithm is achieved by the convention that any excess frequency enters the diagram between the end of the propagator and the (amputated) leg of the vertex. Then, the MWL update only affects the value of a propagator, whereas the MWV and MWH moves only change the adjacent vertex. The resulting acceptance ratios are then simply
\begin{align}
\accratio_{MWL} &= \tilde{G}' / \tilde{G}, & \accratio_{MWV} &= \gamma' / \gamma = \accratio_{MWH} ,
\end{align}
where $\tilde{G} = \tilde{G}_{\kv \nu \sigma}$ and $\tilde{G}' = \tilde{G}_{\kv \pm \vec{\delta}, \nu \pm \Delta, \sigma}$ denote the propagators with old and new 4-momentum indices, respectively, and correspondingly for the vertex.

\begin{figure}[htp!]
    \centering
    \includegraphics[width=\columnwidth]{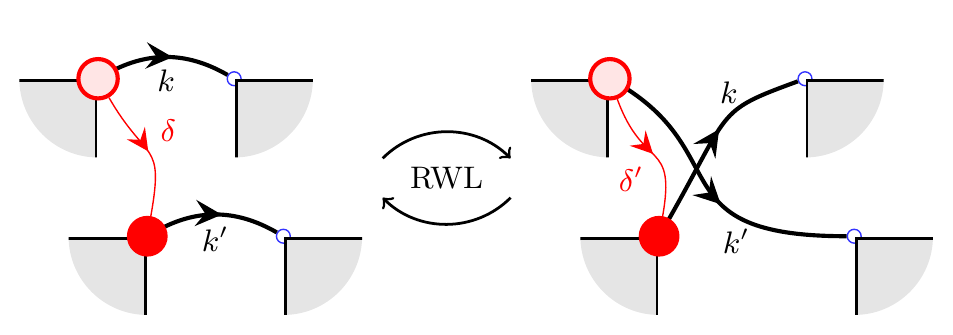}
    \caption{The RWL update reconnects the lines adjacent to the worms. In this example, the new excess 4-momentum is $\delta'=\delta+k-k'$.}
    \label{fig:rwl}
\end{figure}

\paragraph{Reconnect Worm Lines (RWL)}
Diagram topology is changed by reconnecting a pair of propagators, i.e. swapping the locations where the lines end (or where they start). Due to spin and 4-momentum conservation, this is only possible if the worms exist and are both at the end (or both at the start) of different propagators with equal spin indices. Otherwise the update is trivially rejected.
The move does not change the variables associated with the reconnected lines, only the excess 4-momentum $\delta$ associated with the worms changes by the 4-momentum difference $k-k'$ between the two lines. Furthermore, the update may change the number of fermion loops in the diagram. In fact, a moment of thought reveals that it will either merge two loops into one (if both lines were part of distinct loops before) or split one loop into two (if the lines belonged to the same loop). Therefore the total number of fermion loops always changes by $\pm 1$ and hence the diagram's complex sign is multiplied by the factor -1.
In conclusion, the acceptance ratio of this self-conjugate update is
\begin{align}
\accratio_{RWL} = -C_W(n, \delta')/C_W(n, \delta) .
\end{align}
With the above choice for $C_W$, $|\accratio_{RWL}|=1$ unless the new excess frequency would be too large and $P_{\Delta'}=0=C_W(n, \delta')$.

\begin{figure}[htp!]
    \centering
    \includegraphics[width=\columnwidth]{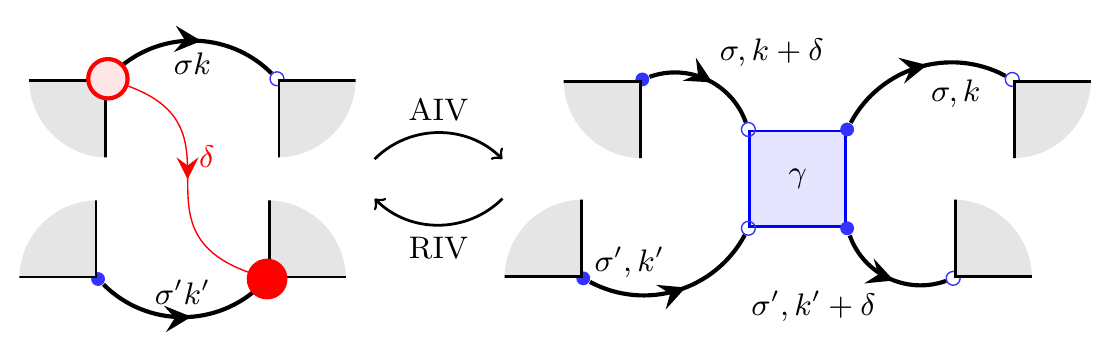}
    \caption{The AIV update replaces the worms by a new interaction vertex $\gamma$. If $\sigma \neq \sigma'$, we may alternatively create a vertex with different spin configuration by connecting the $(\sigma,k)$-line to the lower right and the $(\sigma', k'+\delta)$-line to the upper right corner. The converse move is called RIV.}
    \label{fig:aivriw}
\end{figure}
\paragraph{Add/Remove Interaction Vertex (AIV/RIV)}
The most complex pair of updates changes the diagram order. The AIV update is only possible in the worm sector, if the worms are adjacent to different propagator lines, and if the diagram order $n$ is smaller than the maximum allowed order $N_*$. In this case a new interaction vertex can be inserted into the propagators adjacent to $\ira$ and $\mas$. The new vertex carries the worms' excess 4-momentum $\delta$ from one propagator to the other, such that the worms can be deleted and we are left with a physical diagram of order $n+1$.
If both affected propagators carry the same spin index $\sigma = \sigma'$, we insert the vertex's upper corners into $\mas$'s line and the lower corners into $\ira$'s line, such that the configuration resulting from the move is completely determined by the worms' positions and 4-momentum and the proposal probability is one. 
Otherwise, one of the two spin configurations $\gamma^{\sigma \sigma' \sigma' \sigma}$ or $\gamma^{\sigma \sigma' \sigma \sigma'}$ needs to be chosen. We do so with equal probabilities, so that in general the proposal probability is $\pprop_{AIV}=1/(2 - \delta_{\sigma \sigma'})$. Note that, if $\sigma \neq \sigma'$, the spin configuration of the created vertex uniquely determines what corner each line can be connected to and, if the spin exchange configuration is chosen, the lines are effectively reconnected like in the RWL update such that the number of fermion loops changes by $\pm 1$ and the diagram's complex sign needs to be multiplied by -1.
A final subtlety concerns the case when one of the worms is adjacent to the measuring propagator.
For definiteness we use the convention that, if the vertex is inserted into the measuring propagator, the dummy line will become an outgoing line of the new vertex.

The converse RIV update is trivially rejected if the configuration is already in the worm sector or its order is $n < 2$. Otherwise, a random interaction vertex is chosen and deleted. Since the converse update never creates a vertex at the end of the measuring propagator, we must not remove the vertex at the end of the dummy line and can only choose from $n-1$ vertices. 
The deletion of a vertex results in four dangling propagators. If all carry the same spin index, we merge those that were connected to the vertex's upper corners into a single propagator and correspondingly for the lower ones. If there are different spin indices, the lines with equal spin must be merged and a sign flip due to changed fermion loop count occurs if an upper line merges with a lower line.
Since the merged propagators were typically carrying different 4-momenta, momentum conservation will be broken at one of their ends. This mismatch is accounted for by creating a worm at one end of each propagator. Whether the worm is put at the start or end of a line is arbitrary and we choose randomly between the $2 \cdot 2=4$ possible configurations, so the overall proposal probability is $\pprop_{RIV} = 1/[4 (n-1)]$.

Taking into account that the diagrammatic rules associate with each diagram order a factor $-T/(2 \pi)^d$, we arrive at an acceptance ratio
\begin{align}
R_{AIV} &= \frac{1/4 n}{1/(2 - \delta_{\sigma \sigma'})} \cdot \frac{-s_F T \gamma' (\prod_{i=1}^4 \tilde{G}_i')  /(2 \pi)^d}{(\prod_{i=1}^2 \tilde{G}_i) C_W(n,\delta)}
\nonumber\\&
= - \frac{s_F (2 - \delta_{\sigma \sigma'}) T \gamma' (\prod_{i=1}^4 \tilde{G}_i')}{4 (2 \pi)^d n (\prod_{i=1}^2 \tilde{G}_i) C_W(n, \delta)} ,
\end{align}
where $s_F=-1$ if the updates change the number of fermion loops and +1 otherwise, $n$ is the diagram order before the AIV update, the primed quantities denote the vertex and propagators inserted by AIV, and the non-primed $\tilde{G}_i$ the replaced propagators.

\paragraph{Switch To/From Normalization sector (STN/SFN)}
As the names suggest, these updates switch between the normalization and physical sectors. For simplicity, they are only allowed for Hartree-like first-order diagrams, \ie a single vertex whose upper (or lower) corners are connected by a dual propagator and the lower (or upper) corners by the measuring propagator. If the vertex is a physical interaction vertex $\gamma$, the STN update replaces it by a dummy vertex with unit weight. If, on the other hand, the old configuration was the normalization diagram, the SFN update replaces the dummy vertex by a physical interaction.
Furthermore, we allow for a change of spin indices during this update. This is crucial in ensuring that diagrams with all possible spin indices are sampled, for example both configurations where all propagators carry the same spin and the ones with different spin indices.
Specifically, when going to the normalization sector, we randomly choose the spin assigned to the upper corners of the dummy vertex $\sigma = \uparrow$ or $\downarrow$ and set the spin on the lower corners to $-\sigma$.
When leaving the normalization sector, however, we set the spin indices at the upper and lower corners, respectively, to independent random values.
We therefore have proposal probabilities $\pprop_{STN} = 1/2$ and $\pprop_{SFN} = 1/4$ and are left with the acceptance ratio
\begin{align}
R_{STN} = \frac{\tilde{G}'}{2 \gamma \tilde{G}} .
\end{align}
Here we have included the ratio of propagators before and after the update in order to allow for the possibility of $\tilde{G}$ depending on its spin index. In the paramagnetic phase they will generally be the same and $R_{STN} = 1/(2 \gamma)$.

\paragraph{Swap Measuring Propagator (SMP)}
This update is not necessary for ergodicity, but it is simple and improves the efficiency of sampling different spin and 4-momentum indices of the dual self-energy. It randomly chooses any propagator in the diagram and makes that line the measuring propagator while converting the previous measuring propagator to a regular propagator line. Since the proposal probabilities for a move and its inverse are identical, the acceptance ratio is just given by the ratio of the changed propagators' values.

\subsubsection{Measurements}

After each update, a measurement can be performed according to the standard DiagMC protocol:
If the current configuration is the normalization diagram, the normalization counter is incremented and the measurement is finished.
If the current diagram is physical, \ie not in the worm sector and not reducible, the momentum, frequency and spin indices of the measuring propagator are retrieved, projected onto a suitable basis (like a momentum grid or lattice harmonics basis as detailed below), and the diagram's complex sign is added to the accumulator(s) for the corresponding basis function(s).
If an unphysical diagram other than the normalization one is encountered, no action is required.

As usual in DiagMC, one-particle irreducibility of the diagram can be efficiently checked by taking advantage of momentum conservation: If the diagram is reducible, there is at least one other line with exactly the same 4-momentum index as the measuring propagator. Since in an irreducible topology two lines will almost never have the same 4-momentum, we can treat any such configuration as reducible and hence unphysical. Such a check for the presence of a line with a given momentum is much more efficient than a global topology inspection. In particular, if a suitable hash table is maintained by the updates, the check can be performed in constant time. 

Stochastic error bars on our results are produced by a jackknife analysis over $M$ independent runs, with typically $M \geq 96$. Each run is first equilibrated by performing a number of updates without measurements. In all the cases presented in this paper, autocorrelation times are several orders of magnitude smaller than the $10^7$ thermalization steps we typically perform.

\subsubsection{Optimizations} \label{sec:optim}

Here we mention a few optimizations that greatly reduce the numerical effort required for a given simulation. Some of these are well-known in the dual-fermion or DiagMC communities, respectively, whereas others are specific to the DiagMC sampling of DF diagrams and have not been described before.

\paragraph{Storage of impurity functions}
In practical calculations, the frequency-dependent quantities $\gamma_{\nu \nu' \omega}, \Delta_\nu, g_\nu$ can only be stored up to a finite cutoff frequency. However, we strongly mitigate cutoff effects by replacing propagators beyond the cutoff frequency by their asymptotic high-frequency tails (up to fourth order in $1/\nu$) and vertices by their high-frequency limit $\gamma^{\sigma_1 \sigma_2 \sigma_3 \sigma_4} \to U (\delta_{\sigma_1,-\sigma_2} \delta_{\sigma_2,\sigma_3} - \delta_{\sigma_1,\sigma_2} \delta_{\sigma_2,-\sigma_3})$. For the temperatures considered in this study all cutoff frequencies could be chosen large enough to have no practical relevance.~\footnote{
    A more careful treatment of the vertex's structure at high frequencies~\cite{Wentzell2016} may be useful when pushing the technique to low temperatures.
}

Furthermore, the use of symmetries allows for a significant reduction in memory requirements for storage of the vertex.
In the paramagnetic phase only two spin configurations need to be stored explicitly, \eg $\gamma^{\up \up \up \up}$ and $\gamma^{\up \up \down \down}$. The other non-vanishing configurations can be reconstructed from these as
\begin{align}
\gamma^{\down \down \down \down}  &= \gamma^{\up \up \up \up}, \\
\gamma^{\down \down \up \up} &= \gamma^{\up \up \down \down}, \\
\gamma^{\up \down \down \up} &= \gamma^{\down \up \up \down} = \gamma^{\up \up \up \up} - \gamma^{\up \up \down \down} .
\end{align}
Last but not least, values for negative bosonic frequencies $\omega < 0$ are related to values at $\omega > 0$ by time reversal symmetry
\begin{align} 
\gamma_{\nu \nu' \omega} &= \gamma_{-\nu, -\nu', -\omega}^* 
\end{align}
and do hence not need to be stored explicitly.

\paragraph{Disconnected diagrams}
The RIV and RWL updates as described above may generate a disconnected diagram by removing the only vertex or reconnecting the only two propagators connecting two subgraphs. These disconnected diagrams will be in the worm sector with one worm on each subgraph, so it is impossible to create a disconnected diagram in the physical sector by moving the worms to the same location and invoking the DWO update. For this reason, measurements do not need to explicitly check for disconnected diagrams.
Creation of unphysical disconnected diagrams can be suppressed altogether by setting the worm weight for vanishing excess momentum $C_W(\delta=0)=0$. Due to momentum conservation, zero is the only allowed value for $\delta$ when there is no path between the two worms. If there is such a path, on the other hand, the diagram is connected and $\delta=0$ will almost never appear when continuous momenta are sampled, so that such events have a vanishing contribution to overall statistics.

\paragraph{Measurement in symmetry-adapted basis} \label{sec:kbasis}
Sampling of a momentum-dependent quantity like $\tilde{\Sigma}_{\kv}$ constitutes a classic example of a bias-variance tradeoff: Reconstruction of the $\kv$-dependence with high resolution, \eg by measuring on a fine $\kv$-grid, implies large stochastic errors because only few samples effectively contribute to each basis function. The stochastic errors can be reduced by choosing a lower resolution and effectively averaging over more samples, but this increases the bias of projecting the continuous function on a smaller number of basis functions.

Use of spin, time-reversal, and point group symmetries allows to reduce the number of independent basis functions and hence the variance without introducing an unphysical bias. Additionally, the self-energy is known to decay in real space, so capturing its value on short to intermediate distances is in general more important. For these reasons, we measure $\tilde{\Sigma}_\kv$ in a lattice-harmonics basis, \ie measurements are projected on symmetry-adapted combinations of $\cos n_i k_i$ with integers $n_i = 0, \dots, n_{\text{max}}$. As these integers directly correspond to distances in real space, a cutoff to $n_i \leq n_{\mathrm{max}}$ effectively corresponds to a restriction of $\tilde{\Sigma}_{\vec r}$ to a finite area in real space.
We stress that the complete simulation is performed for an infinite system with continuous momenta and only the measurements project on a finite basis. Variation of $n_{\mathrm{max}}$ during the evaluation allows for a simple \emph{a posteriori}-check that the chosen basis is large enough and does not introduce any significant artifacts.
For details on the generation of suitable basis functions for a given lattice the reader is referred to Ref.~\onlinecite{gukelberger2015diss}.

\paragraph{Frequency range of the dual self-energy}
The dual self-energy decays quickly with frequency because the leading-order tails of the lattice and impurity self-energies agree. Therefore it is often beneficial to restrict the sampling process for $\tilde{\Sigma}_\nu$ to a finite window $\nu \in [-\nu_c, \nu_c]$ by rejecting any update that would set the measuring line's frequency to a value outside this window, and checking a posteriori that the window was large enough.

\paragraph{Local propagators}
A large fraction of valid diagrams contains at least one local propagator, \ie a propagator line whose both ends are connected to the same interaction vertex. If this line represents the bare dual propagator $\tilde{G}^{(0)}$ based on the DMFT hybridization $\Delta^{\textrm{DMFT}}$, we know it to be purely non-local per the discussion below Eq.~\eqref{eq:sigma-from-sigmad}. In other words, the line's momentum integral will vanish and all diagrams with local propagators do not contribute to the final result.
In order to avoid spending a major part of the simulation time on the sampling of these irrelevant diagrams, the generation of configurations with local propagators should be suppressed altogether.
In fact it is straightforward to add a check to the topology-changing updates RWL and RIV that rejects any move that would create a local propagator.
Note that this optimization cannot be employed when the dual-fermion expansion is based on a different hybridization $\Delta \neq \Delta^{\textrm{DMFT}}$ because then the bare dual propagator generally does have a non-vanishing local contribution.

\subsection{Diagrammatic Determinant Monte Carlo}
\label{sec:ddmc}

We briefly describe the DDMC algorithm we use for the benchmarks.
DDMC stochastically sums the weak-coupling expansion series for the finite-temperature partition function $Z$ (for details, see Refs.~\onlinecite{DDMC_njp2006, DDMC_2013}). Expanding $Z$ with respect to the interaction $U$ generates the standard Feynman diagrammatic series: at a given expansion order $p$ there are $(p!)^2$ diagrams corresponding to all possible ways of connecting $p$ four-pole interaction vertices by single-particle non-interacting propagators. Given a fixed configuration of $p$ vertices, summing over all the interconnections for each spin component amounts to computing a determinant of a $p \times p$ matrix consisting of non-interacting propagators. This effectively reduces the factorial scaling of computational complexity with the typical diagram order $N$ to $N^3$. Deterministic summation of all the diagrams for a particular vertex configuration by means of calculating determinants is what contrasts DDMC with DiagMC. In DiagMC self-energy diagrams have to be sampled one by one since they cannot be collected into a single determinant. In addition, the particle-hole symmetry of the half-filled Hubbard model on a bipartite lattice~\cite{Micnas1990} allows to change the sign of $U$ by an exact mapping that makes the expansion in terms of such determinants sign-positive. Therefore this expansion can be efficiently summed by a straightforward Metropolis-type scheme, which does not suffer from the fermionic sign problem. A Monte Carlo configuration is a particular arrangement of $p$ vertices on the lattice and in continuous imaginary time, while the updates are sampling over all possible arrangements and different numbers of vertices $p$.   

Since $Z$ is an extensive quantity, one has to confine the simulation volume to a finite lattice of $L \times L$ sites with periodic boundary conditions (phase-twisted boundary conditions can also be used to reduce shell effects \cite{DDMC_2013}). On a finite lattice, the diagrammatic series converges, and thus DDMC produces numerically exact results for a system of finite linear size $L$. An extrapolation with respect to $L \to \infty$ -- based on data obtained for several values of $L$ -- is therefore necessary to claim unbiased results in the thermodynamic limit. The maximal accessible system size is limited by the efficiency of the algorithm, the bottleneck of which is the calculation of determinants for each diagram order $p$. Thus the computational complexity of DDMC on a square lattice scales as~\cite{DDMC_2013} $[U*L^2/T]^3$ and systems of sizes up to $L \sim 20$ can be reliably addressed in the range of parameters studied here. This is typically sufficient for obtaining thermodynamic-limit results with acceptable error bars for the purpose of benchmarking the approximate methods studied here.

\begin{figure}[tb]
    \centering
    \includegraphics[width=\columnwidth]{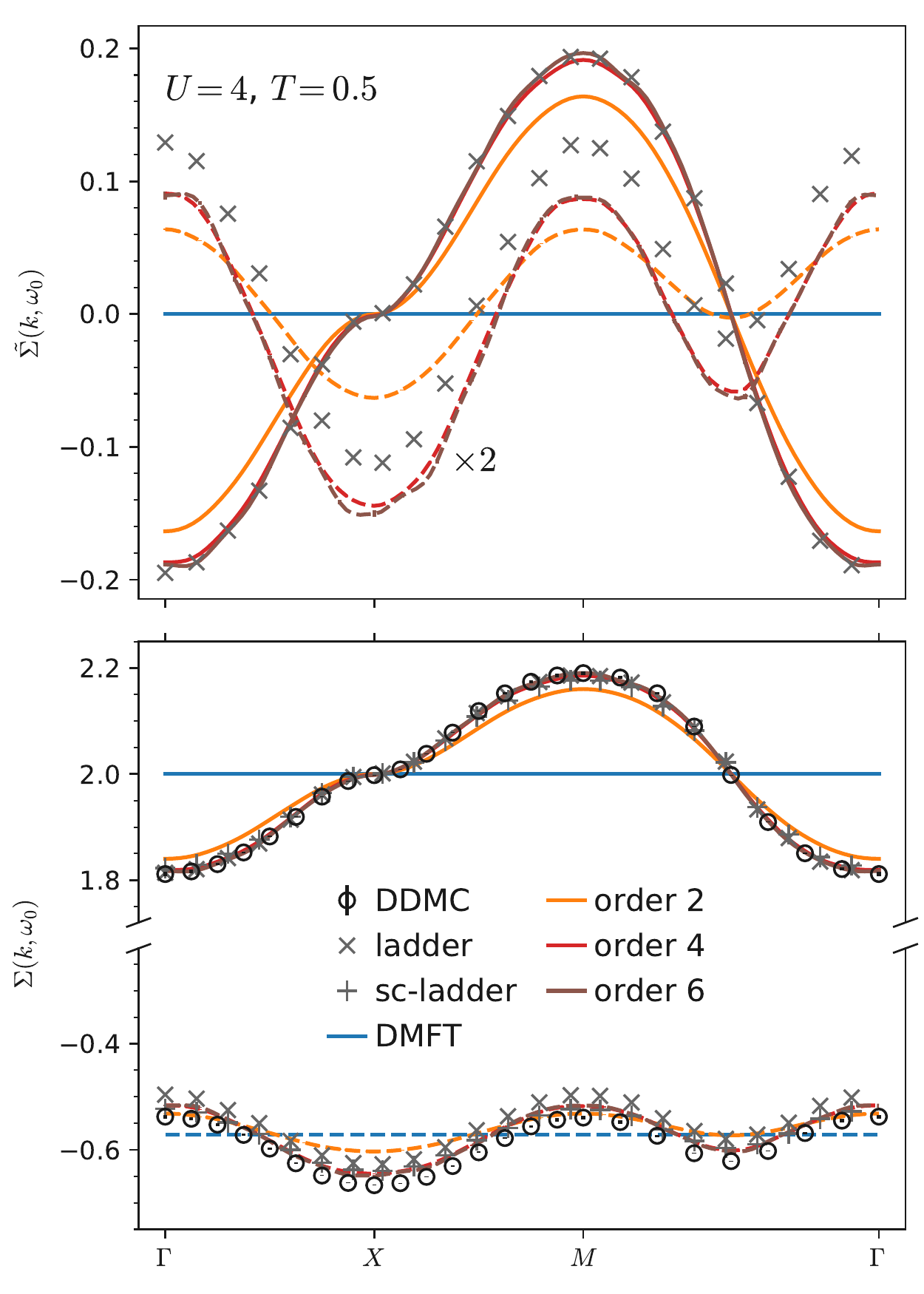}
    \caption{Dual (top) and lattice-fermion (bottom) self-energy at the lowest Matsubara frequency along the high-symmetry lines of the Brillouin zone for moderate interaction $U=4$ and temperature $T=0.5$ at half filling $n=1$. The imaginary part of $\tilde{\Sigma}$ has been multiplied by a factor of two for better visibility.
    Different colors correspond to diagram order cutoffs $N_*=$ 1 (DMFT), 2, 4, 6. Grey crosses indicate results from the one-shot ($\times$) and self-consistent ($+$) ladder approximation, circles the DDMC benchmark. Real (imaginary) parts are shown with solid (dashed) lines. Stochastic errors on DDMC and DiagMC@DF data are displayed as vertical bars, but often indiscernible because they are smaller than the line width.}
    \label{fig:u4_b2_n1-sigma}
\end{figure}

\begin{figure*}[tp!]
    \centering
    \includegraphics[width=\columnwidth]{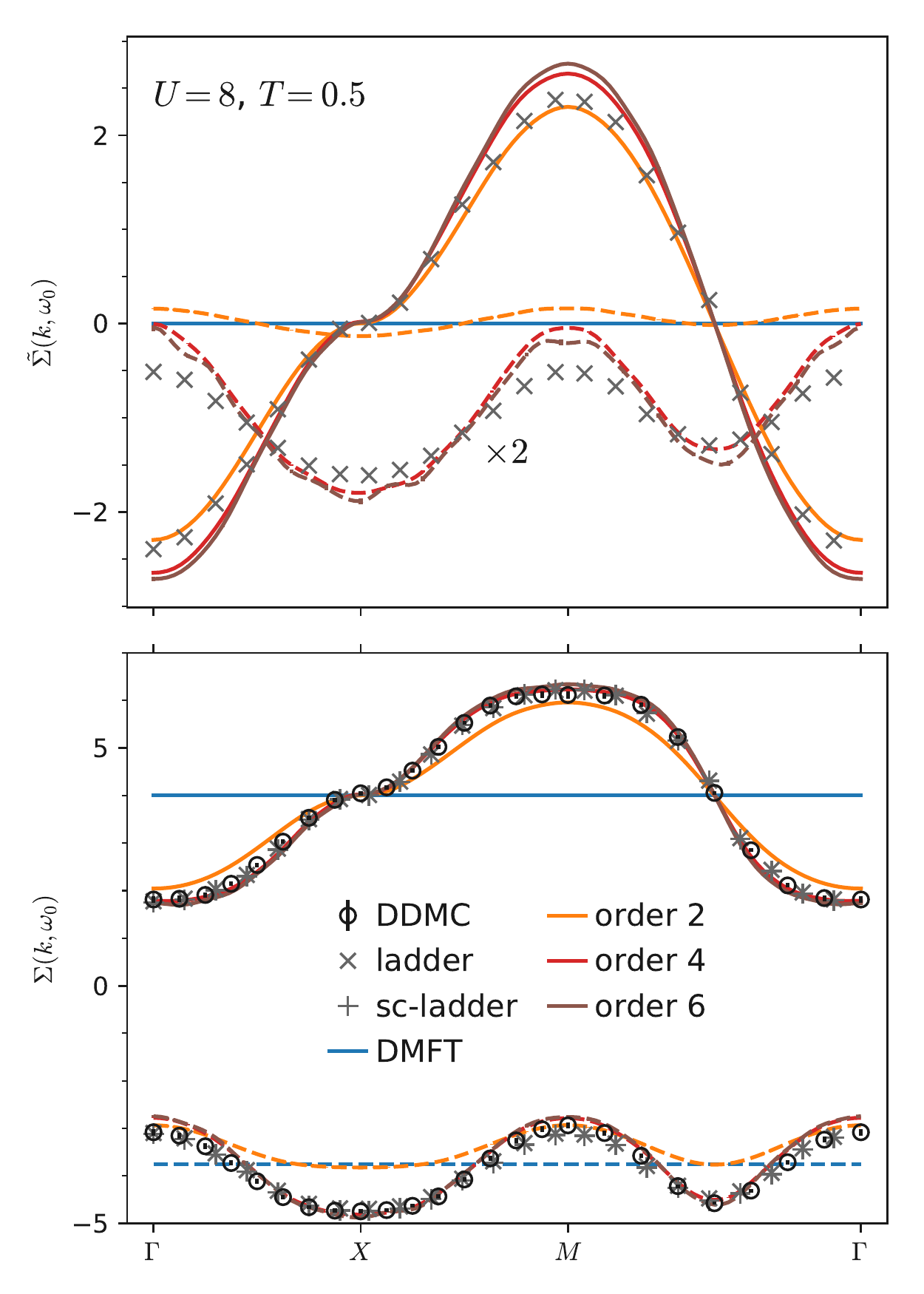}
    \includegraphics[width=\columnwidth]{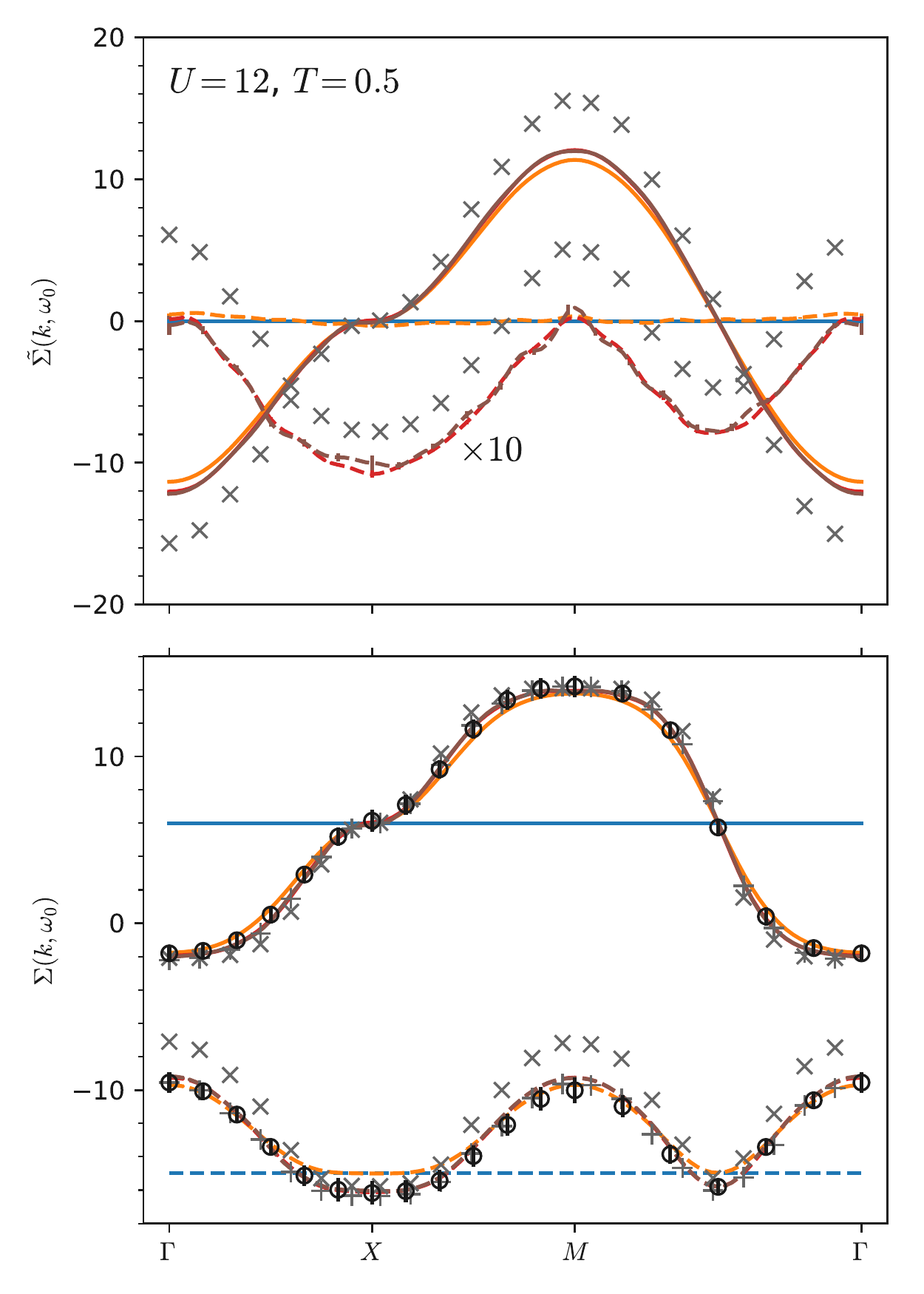}
    \caption{Dual (top) and lattice-fermion (bottom) self-energy at the lowest Matsubara frequency along the high-symmetry lines of the Brillouin zone at half filling and $T=0.5$ for stronger interactions $U=8$ (left) and $U=12$ (right). For better visibility, the imaginary parts of the dual self-energies have been magnified by factors of 2 and 10, respectively.
    See Fig.~\ref{fig:u4_b2_n1-sigma} for an explanation of the different lines and symbols.}
    \label{fig:u8n12_b2_n1-sigma}
\end{figure*}

\section{Results} \label{sec:results}

In the following we present detailed benchmarks of the technique. For different representative parameter regimes, we evaluate the convergence properties of the dual-fermion series including all topologies with two-particle interactions. In particular we compute dual and lattice self-energies for successively higher values of the diagram-order cutoff to study the convergence of the diagrammatic series  and compare these results to the frequently used second-order and ladder approximations to the DF series. By the latter we mean the ladder approximation in the symmetrized theory, which contains contributions from the horizontal and vertical particle-hole channels. We note that these approximations employ a self-consistent renormalization of the dual propagators while we sample diagrams with bare propagators in DiagMC@DF. We further refer to ladder DF results for which the hybridization has been adapted to fulfill the self-consistency condition $\sum_{\kv}\tilde{G}_{\kv\nu\sigma}=0$ in terms of the renormalized propagators as self-consistent results (see Sec.~\ref{sec:df}).
Except for the comparison in Sec.~\ref{sec:hybchoice}, all our DiagMC@DF results are based on the DMFT hybridization $\Delta^{\textrm{DMFT}}$.

At half filling (Sec.~\ref{sec:halffilling}), comparisons to numerically exact DDMC simulations allow us to validate converged results and assess the effects of neglecting higher-order vertices.
Having obtained a good understanding of the DF series' behavior in different interaction and temperature regimes we then turn to densities away from half filling (Sec.~\ref{sec:doped}), where numerically exact techniques are lacking in general.
Finally, we address some technical questions regarding the DF approach and possible improvements in Secs.~\ref{sec:topologies} - \ref{sec:hybchoice}.

The reader can find raw data from all our simulations in the supplemental material of this manuscript.~\cite{suppmat} This includes the hybridization function used as input for the DiagMC@DF calculations and lattice self-energies obtained by DiagMC@DF as well as the ladder DF approximations, DDMC and DiagMC benchmarks.

\subsection{Dual-fermion expansion at half filling} \label{sec:halffilling}

\begin{figure}[htb]
    \centering
    \includegraphics[width=\columnwidth]{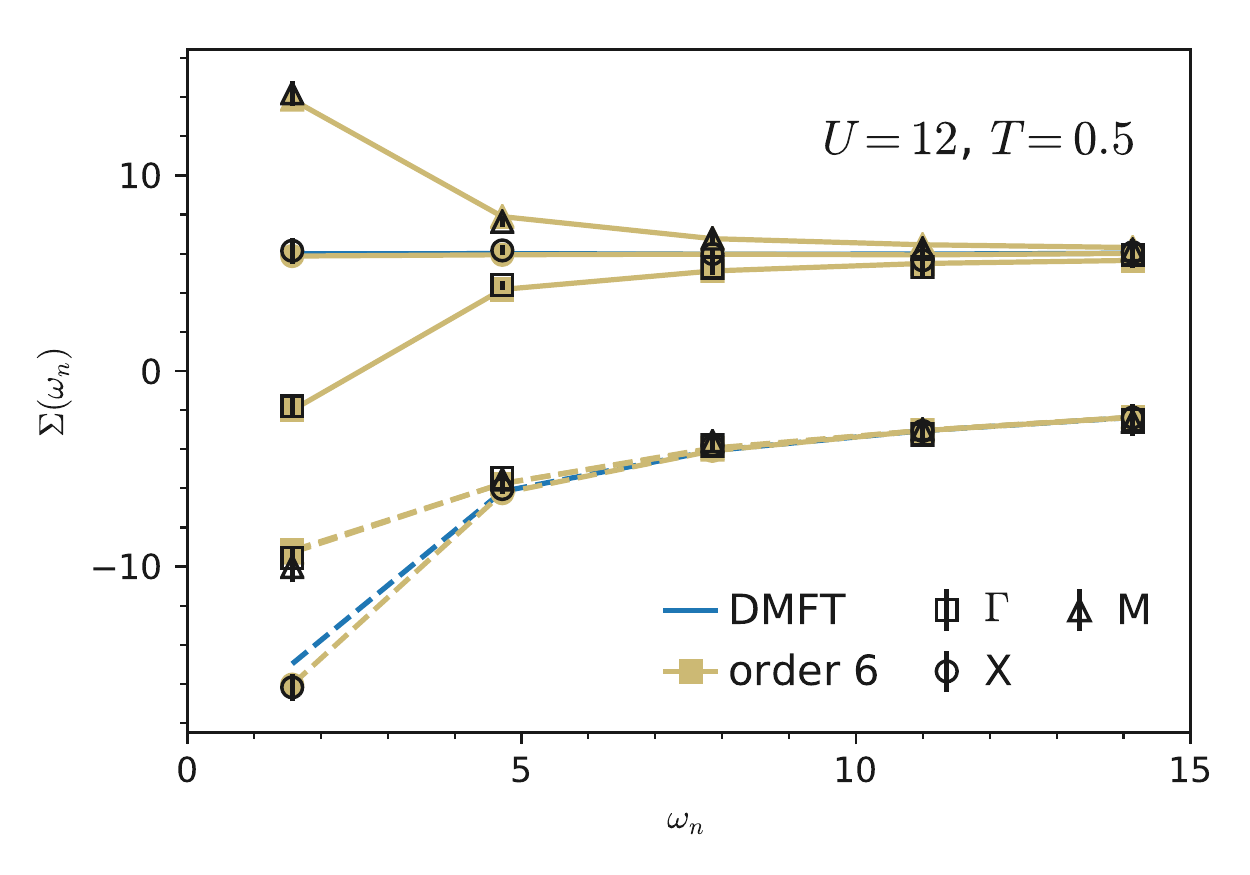}
    \caption{Lattice self-energy for $U=12$, $T=0.5$, $n=1$ at the high-symmetry points $\Gamma(0,0)$, X$(\pi,0)$, and M$(\pi,\pi)$ plotted versus Matsubara frequency $\omega_n$. Shown are the 6th-order DiagMC@DF result (yellow filled symbols) and the DDMC benchmark (open black symbols) as well as the momentum-independent DMFT data (blue lines). Symbols referring to real and imaginary parts are connected by solid and dashed lines, respectively.}
    \label{fig:u12_b2_n1-sigma_vs_w}
\end{figure}

We first concentrate on three representative values of the interaction, namely $U=4$ (Fig.~\ref{fig:u4_b2_n1-sigma}), 8, and 12 (Fig.~ \ref{fig:u8n12_b2_n1-sigma}) at moderate temperature $T=0.5$. The first interaction value is on the weak-coupling side of the phase diagram and smaller than the critical value for the finite temperature Mott transition within DMFT ($U_{c}=9.35$) and cellular DMFT ($U_{c}=6.05$).~\cite{Park2008}
The second case corresponds to the strong-correlation regime with possible relevance to the cuprate superconductors and the last one to the strong-interaction regime.
Figures \ref{fig:u4_b2_n1-sigma} and \ref{fig:u8n12_b2_n1-sigma} show the momentum dependence of dual and lattice self-energies at the lowest Matsubara frequency $\omega_0 = i \pi T$.
In all cases rapid convergence with diagram order to the exact solution is apparent: While the first-order result corresponds to the momentum-independent DMFT self-energy, the second-order diagrams already contribute a major part of the correct momentum variation and fourth-order results are essentially indistinguishable from the exact solution.
For comparison, we also indicate results from the popular DF ladder approximation: Grey $\times$ symbols correspond to the one-shot ladder approximation whereas $+$ symbols represent lattice self-energies resulting from self-consistent ladder calculations.
We do not display dual self-energies for the self-consistent calculations because dual self-energies cannot be meaningfully compared when they are based on different hybridization functions.
While differences between the one-shot ladder approximation and the converged dual self-energy prove that non-ladder diagrams do give a non-zero contribution, a self-consistent adjustment of the hybridization function is apparently enough to reproduce very accurate results for the lattice self-energy at this temperature, which confirms what has been found in earlier benchmark comparisons.~\cite{Leblanc2015} 
We do not explicitly show results from calculations that iterate the second-order approximation to self-consistency; we found that this scheme typically underestimates the self-energy and is inferior to both the one-shot realization of the second-order approximation and the self-consistent ladder approximation.

The impressive agreement between the converged results for the lattice self-energy and DDMC reference data further shows that in the considered cases the effect of higher-order interaction vertices must be essentially negligible. We discuss this point below in more detail.
We note, however, that the efficiency of the DDMC and DiagMC@DF algorithms at high temperature and moderate coupling allows us to resolve small differences in the imaginary part of the self-energy throughout the Brillouin zone for $U=4$ and at some $k$ points for $U=8$. Since it would be impossible to resolve such small differences at lower temperatures or stronger interactions, where error bars are significantly larger, we consider converged self-energies with deviations from the benchmark $|\Delta \Sigma|/|\Sigma| < 1\%$ as exact. One should however keep in mind that there is no reason in general why contributions from higher-order vertices should vanish identically and benchmarks with sufficiently high accuracy will be able to resolve the effect of neglecting a small but nonzero contribution.

Figure~\ref{fig:u12_b2_n1-sigma_vs_w} shows the frequency dependence of the lattice self-energy for the example of $U=12$. It is obvious that the nonlocal corrections quickly decay with frequency because the high-frequency tail of the self-energy is local and accurately captured by DMFT. The real and imaginary parts of the dual self-energy decay with $\mathcal{O}(1/\omega_n^2)$ and $\mathcal{O}(1/\omega_n^3)$, respectively, so that the DF result for the lattice self-energy inherits the exact high-frequency asymptotics from the DMFT solution. This is an advantage over the \dga{} and 1PI approaches, where the correct high-frequency behavior must be enforced by additional correction factors, unless the self-energy is computed in the parquet approximation.
In the following we will concentrate on the lowest Matsubara frequency, where the non-local corrections are largest.

\begin{figure}[htp!]
    \centering
    \includegraphics[width=\columnwidth]{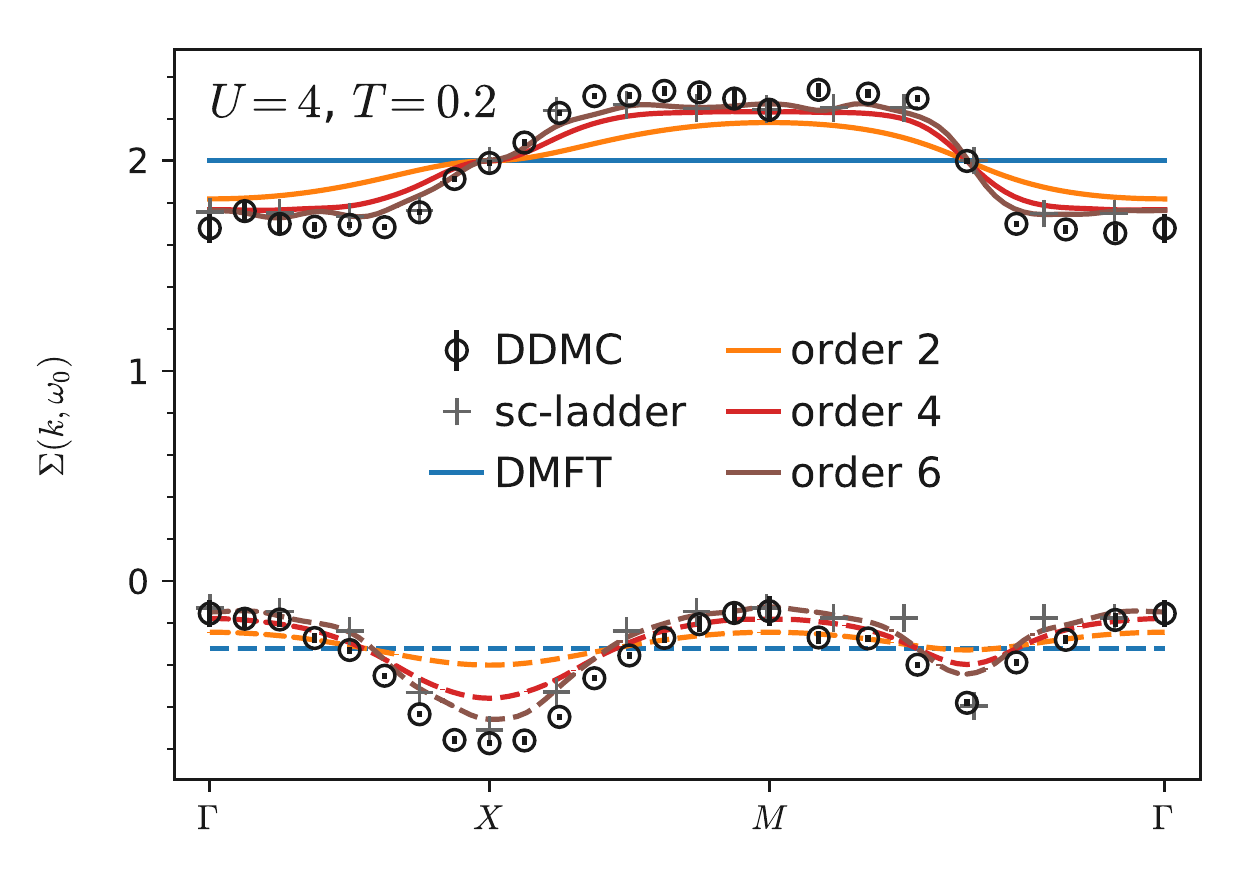}
    \includegraphics[width=\columnwidth]{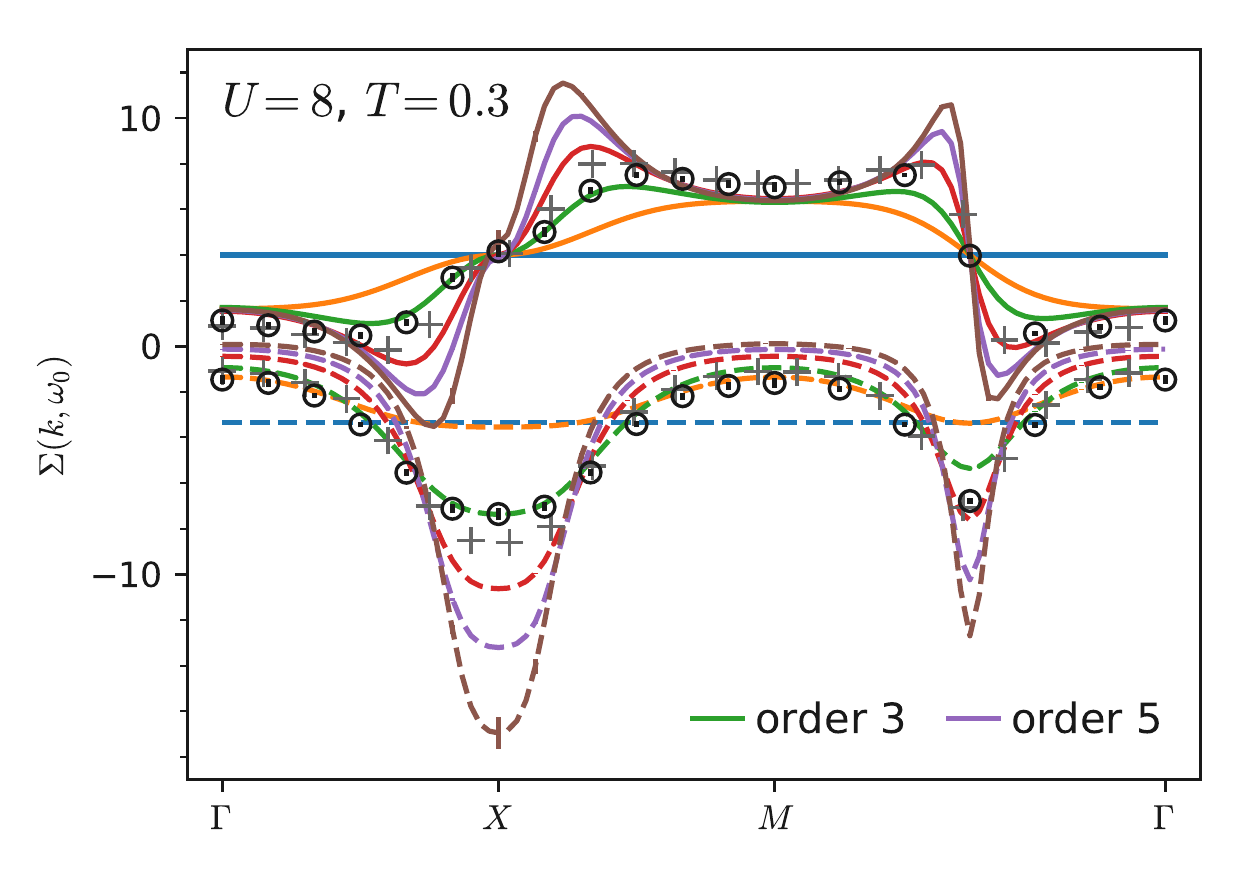}
    \includegraphics[width=\columnwidth]{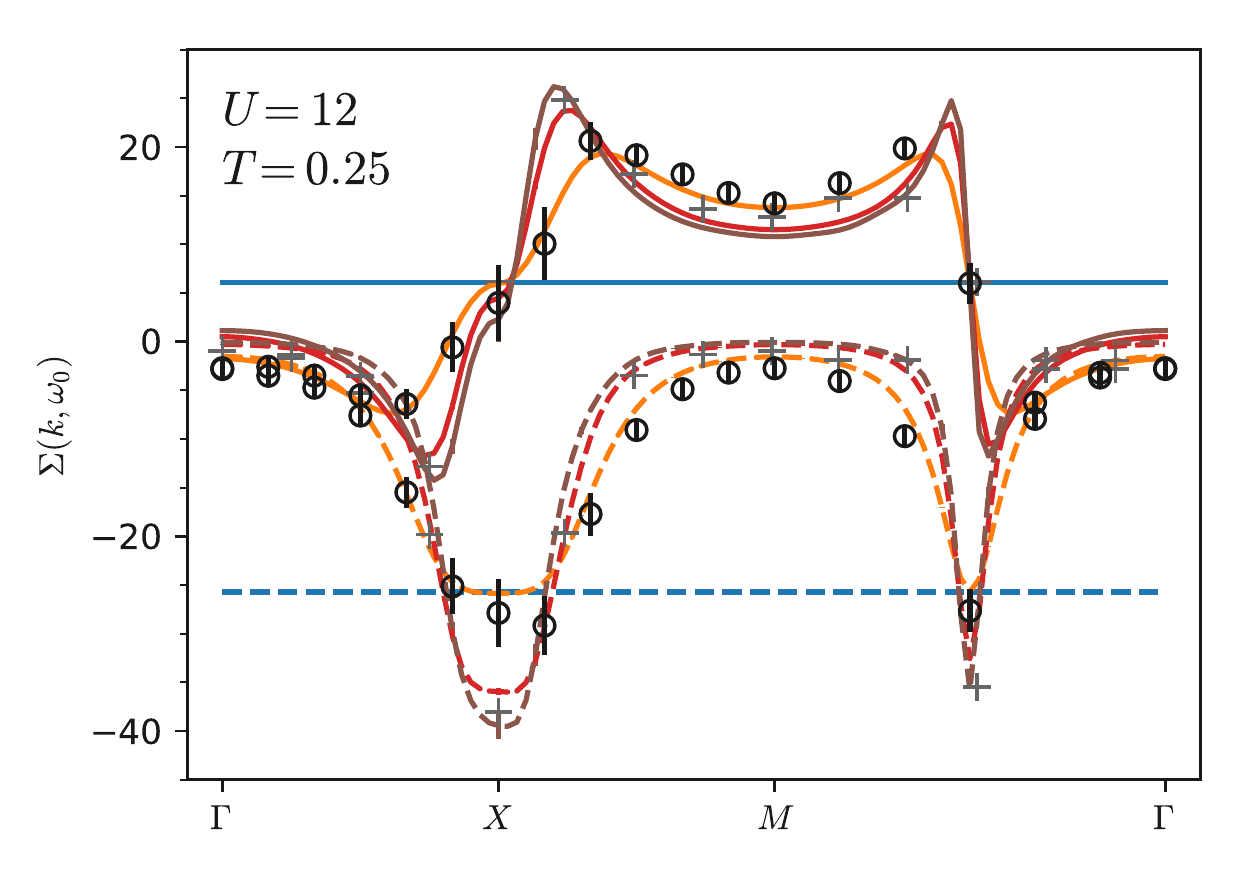}
    \caption{Lattice self-energies at half filling and lower temperatures. The parameters are $U=4$, $T=0.2$ (top), $U=8$, $T=0.3$ (center), and $U=12$, $T=0.25$ (bottom). See Fig.~\ref{fig:u4_b2_n1-sigma} for an explanation of the different lines and symbols.}
    \label{fig:lowT-sigma}
\end{figure}

\begin{figure}[htp!]
    \centering
    \includegraphics[width=\columnwidth]{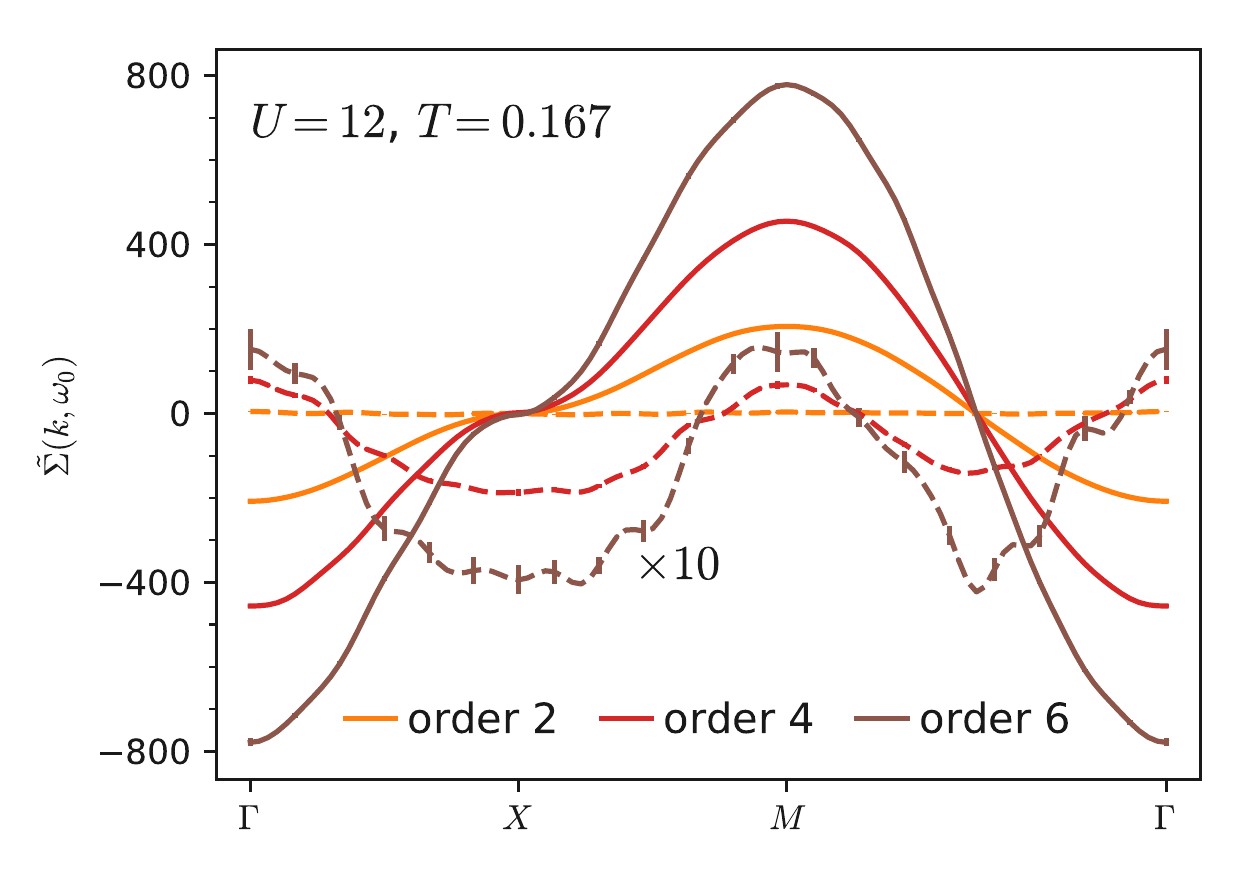}
    \caption{The dual self-energy series at half filling and $U=12$, $T=0.167$ appears strongly divergent. For better visibility, the imaginary part is magnified by a factor of 10. Real (imaginary) parts are shown with solid (dashed) lines. Stochastic errors are displayed as vertical bars, but often indiscernible because they are smaller than the line width.}
    \label{fig:divergent-sigmad}
\end{figure}

At lower temperatures, the DF series converges significantly less quickly as can be seen in the examples of Fig.~\ref{fig:lowT-sigma}, which shows lattice self-energies for the same interactions as before but at lower $T$.
Additional issues are apparent in the different subplots. 
At $U=4$, $T=0.2$ the series seems to converge slowly towards the correct solution. In this case a sign problem makes it computationally very expensive to converge the stochastic DiagMC error bars for higher diagram orders and ultimately prevents us from examining series convergence properties at even lower temperatures. This sign problem is likely connected to the fact that the metallic dual-fermion propagators in Matsubara-frequency space $\tilde{G}(\vec{k}, i \omega_n \to 0)$ acquire large values close to the Fermi surface. It could in principle be resolved by performing the DiagMC sampling in a different representation such as imaginary-time space. The latter is standard practice for conventional DiagMC schemes sampling the weak-coupling expansion,~\cite{VanHoucke2010} where no temperature-dependent sign problem is observed in the Fermi liquid regime. 
Unfortunately, the Fourier transform of the reducible vertex to imaginary times is numerically much less convenient and contains delta functions whenever two or more time arguments coincide. We therefore defer this issue to further studies.

In contrast to the previous cases, at $U=12$, $T=0.25$, the series slowly converges towards a self-energy that differs significantly from the exact solution, whereas at $U=8$, $T=0.3$ there are no signs of convergence at all. The self-consistent ladder result however, in which the diagrams are evaluated using renormalized propagators, is still close to the DDMC result.
Also on the strong-interaction side $U=12$, series divergence can be observed at lower temperatures, as demonstrated in Fig.~\ref{fig:divergent-sigmad}. 
Here the dual self-energy quickly becomes so large that poles appear in the mapping of the dual to the lattice self-energy Eq.~\eqref{eq:sigma-from-sigmad} and no meaningful lattice self-energy can be computed.

\begin{figure}[tb]
    \centering
    \includegraphics[width=\columnwidth]{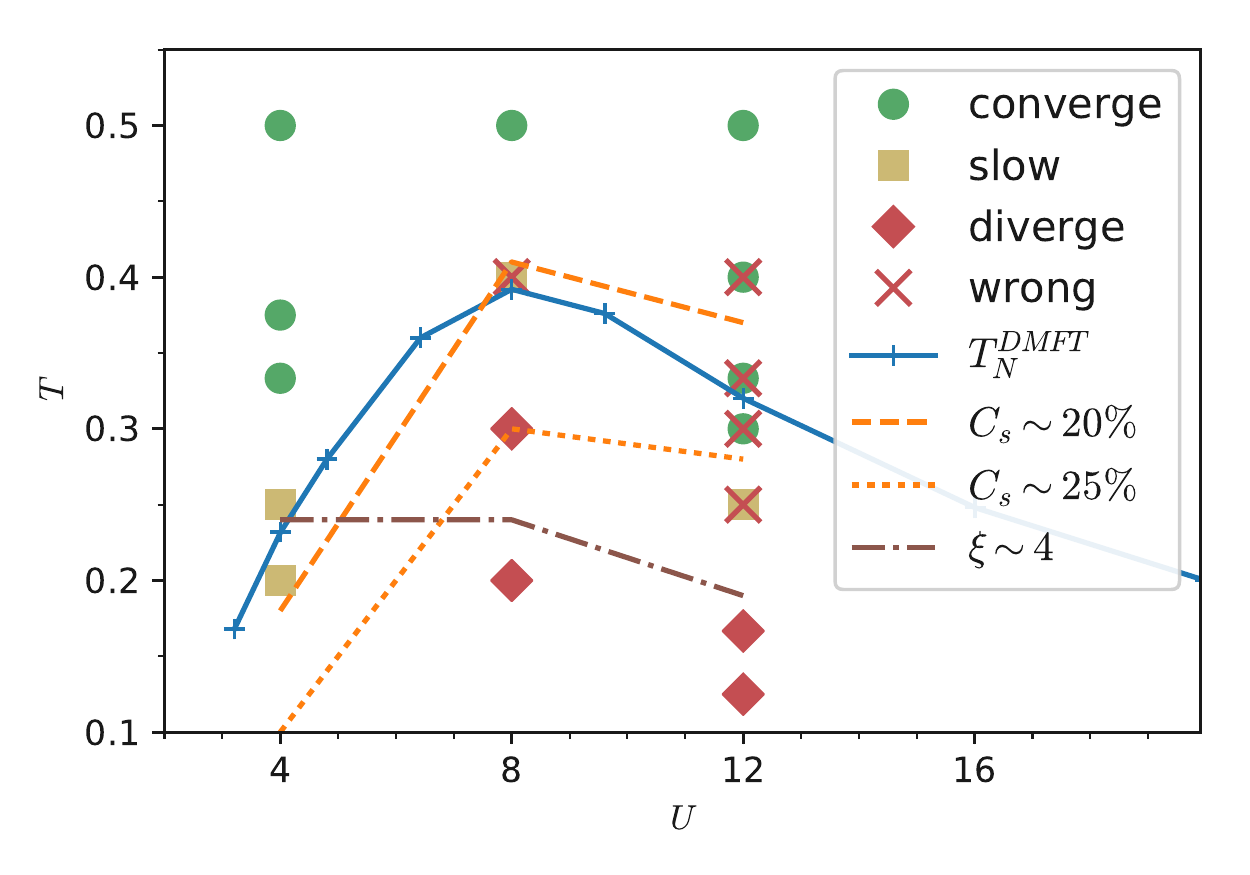}
    \caption{Overview of our DiagMC@DF simulations at half filling (filled symbols). Red diamonds mark parameters where the DF series appears to diverge. Green circles (yellow squares) indicate rapid (slow) series convergence, red crosses convergence to incorrect results. For comparison, the lines show DMFT Néel temperatures from Ref.~\onlinecite{Kunes2011} and temperatures where nearest-neighbor spin correlations $C_s = -4 \langle S^z_i S^z_{i+1} \rangle$ or the correlation length $\xi$ reach a given value (see main text for details).}
    \label{fig:phasediag}
\end{figure}

\subsubsection{Temperature scale where DF expansion breaks down}
\label{sec:tbreakdown}

For all considered interactions, convergence of the DF expansion with diagram order becomes slower at low temperatures and eventually the series even starts to diverge.
Figure~\ref{fig:phasediag} gives an overview of where in the $T-U$ plane these problems start to appear: 
The crossover from fast to slow convergence and further to divergence is visible in the succession of green circles, yellow squares, and red diamonds. We classify a case as slowly convergent when the self-energy appears to converge to a finite result, but the fourth-order result still differs significantly from higher orders, such that an accurate determination of the converged result becomes challenging. Cases marked as divergent show no sign of convergence up to sixth order.
Red crosses indicate cases where the series converges towards a result that differs significantly from the exact solution.
We note that at $U=8$ series divergence sets in at rather high temperature, whereas at $U=12$ there is a comparably broad window $0.25 \lesssim T \lesssim 0.4$ where the series converges to a quantitatively incorrect but qualitatively reasonable solution before the series becomes divergent at even lower temperature.

In all cases the breakdown regime agrees rather well with the temperature range where the system develops significant magnetic correlations.
Physically, both short-range correlations become stronger and the correlation length grows exponentially with $1/T$ after entering into the renormalized classical regime, known to cause a pseudogap phenomenon.~\cite{Vilk:1996,Vilk:1997}
Short-range magnetic order is directly accessible in our DDMC reference simulations via the nearest-neighbor magnetic correlation function $C_s = -4 \langle S^z_i S^z_{i+1} \rangle$ (normalized such that the maximum possible value is unity). In Fig.~\ref{fig:phasediag} we indicate the approximate locations where $C_s$ has grown to 20\% and 25\%, respectively.
As a proxy for the development of quasi-long-range correlations, we consider the system size dependence of the antiferromagnetic structure factor measured in the DDMC simulations: at the $\xi \sim 4$ line $S_{AFM}(L=8)$ starts to differ significantly (i.e. by more than 10\%) from $S_{AFM}(L=12)$, witnessing the presence of non-negligible correlations over at least half the linear extent of the smaller system.

In correspondence to the buildup of magnetic correlations in the physical model, DMFT predicts an antiferromagnetically ordered state below a finite Néel temperature $T^{\mathrm{DMFT}}_N$, in violation of the Mermin-Wagner theorem.
Our DF simulations below $T^{\mathrm{DMFT}}_N$ hence use the metastable paramagnetic solution of DMFT.
Since the impurity construction of DMFT can neither account for short-range singlet formation nor for quasi-long-range magnetic correlations, it is not unexpected that at low temperature the DMFT solution for the half-filled Hubbard model is so far away from the exact solution that the perturbative expansion around the former fails.

Despite the breakdown of the series at low temperature we find that sampling diagrams up to a low cutoff can yield reasonable results.
For $U=0.8$, $T=0.3$ and $U=12$, $T=0.25$ in Fig.~\ref{fig:lowT-sigma} the results for a cutoff of $N_*=3$ are closest to the benchmarks. For $U=4$, $T=0.2$ the result can still be significantly improved by including higher orders, because of the slow convergence of the series. 
This finding is consistent with the fact that the particle-hole ladder diagram series with self-consistently renormalized propagators appears to behave like an asymptotic series at temperatures significantly below $T^{\mathrm{DMFT}}_N$: while at second order the approximation captures the first-order Mott transition at a critical $U_{c}\approx 6.64$ at $T\approx 0.11$ that is close to the value obtained in cluster approaches,~\cite{Hafermann2009} higher orders give unphysical results.~\cite{Hafermanntbp} We note that our DiagMC@DF results cannot be directly compared to these calculations: First, DiagMC is not restricted to ladder-type diagrams and second, the dual Green's functions in the ladder-type approximation of order $n$ contain \emph{all} self-energy insertions constructed from  ladder diagrams up to this order (because they are included via Dyson's equation), while in DiagMC@DF they are limited by the diagram order cutoff.

From Fig.~\ref{fig:phasediag} one is tempted to conclude that the breakdown of the series is directly linked to $T^{\mathrm{DMFT}}_N$. While this is true for the ladder diagram series in terms of bare dual propagators which diverges at $T^{\mathrm{DMFT}}_N$,~\footnote{It can be shown that the DMFT susceptibility and the one calculated in terms of dual fermions are exactly equal as long as no diagrammatic corrections to the self-energy are taken into account.~\cite{Hafermann2009a} In particular, the leading eigenvalues of the Bethe-Salpeter equation in DMFT and in DF cross the value 1 simultaneously. This means that the particle-hole ladder diagram series in terms of bare DF propagators diverges at $T^{\mathrm{DMFT}}_N$.} this need not be the case for DiagMC@DF. Because the algorithm samples all possible topologies, cancellations may occur.
We refer the reader to Ref.~\onlinecite{Gukelberger2015} for an example from conventional (weak-coupling) DiagMC where the bare particle-hole ladder series diverges quickly, whereas the sum over all topologies stays close to the second-order result.

The ladder approximation in terms of self-consistently renormalized propagators, on the other hand, yields a finite result. Here the self-energy insertions contain ladder diagrams up to infinite order that describe the fluctuations that destroy the mean-field long-range order. The result is close to the benchmarks for $U=4$, $T=0.2$ and $U=0.8$, $T=0.3$ (upper and middle panels of Fig.~\ref{fig:lowT-sigma}), despite the fact that in the latter case the series sampled by DiagMC@DF appears to diverge. A bold DiagMC@DF scheme which includes ladder diagrams into the propagators may improve the convergence.
At $U=12$, $T=0.25$ the series however seems to converge to a value which is very close to the ladder result, but differs quantitatively from the exact result.

\begin{figure}[tb]
    \centering
    \includegraphics[width=\columnwidth]{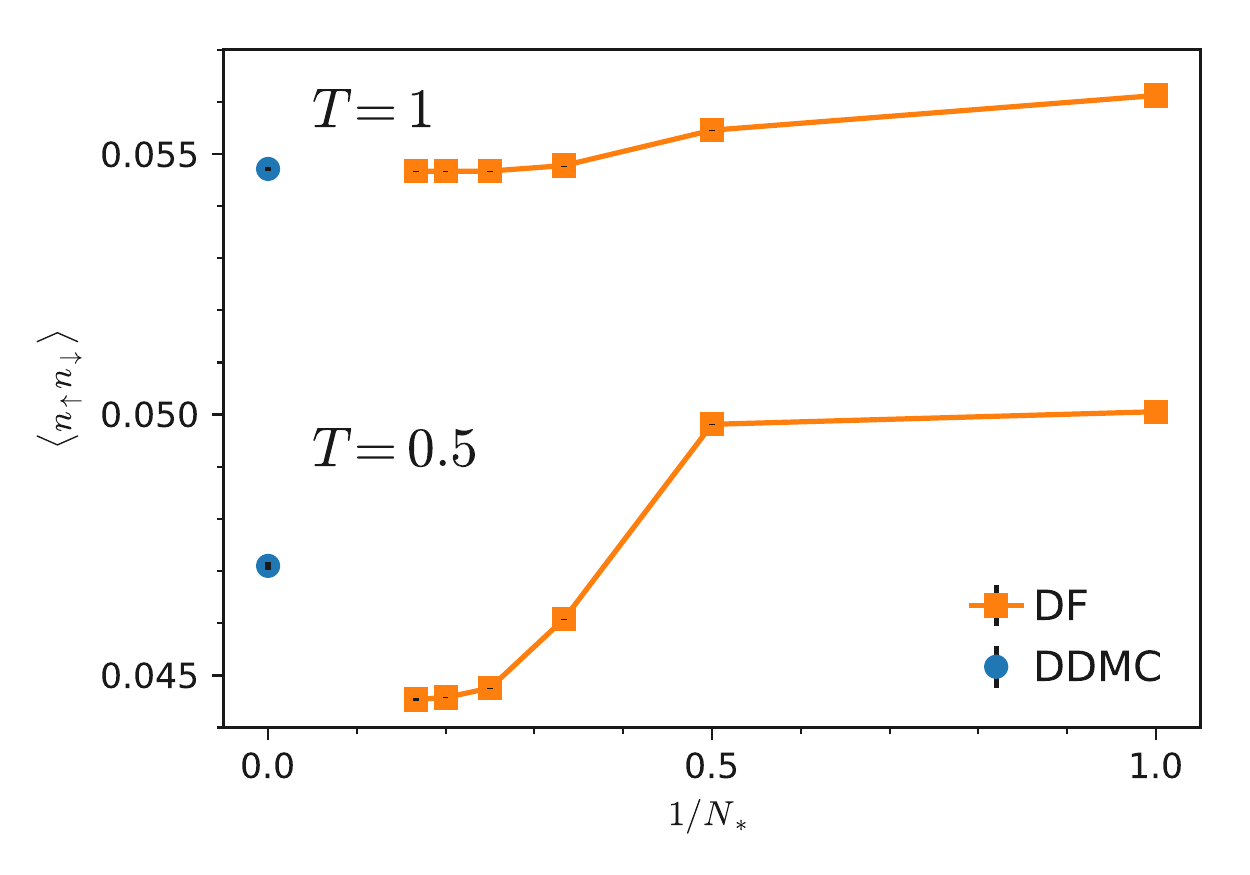}
    \caption{Double occupancy of the half-filled model at $U=8$ and temperatures $T=0.5, 1$. Shown are DF series convergence with diagram order $N_*$ and finite-size extrapolated DDMC results. Diagram order $N_*=1$ corresponds to DMFT.}
    \label{fig:u8-doubleocc}
\end{figure}

\subsubsection{Scalar observables}

The present work concentrates on the non-local correlations, which are completely neglected in DMFT and recovered by the diagrammatic extension. For this reason we mainly investigate the momentum dependence of the self-energy. While in principle an accurate determination of the self-energy implies accurate knowledge of all thermodynamic one-particle observables, one should keep in mind that stochastic and systematic error bars on momentum- and frequency-resolved observables are generally considerably larger than for a scalar observable. In some scalar observables one may therefore observe systematic deviations of the converged DF result from the exact solution already at temperatures where $\Sigma(k)$ still appears perfectly compatible with the benchmarks. This is demonstrated for the double occupancy in Fig.~\ref{fig:u8-doubleocc}: At $T=1$ the DF corrections lead to a clear improvement over the DMFT value and the converged DF and DDMC results match perfectly within small error bars.~\footnote{There is an ambiguity in the definition of the double occupancy within DMFT, as pointed out in Ref.~\onlinecite{VanLoon2016}. The DMFT results quoted here are impurity double occupancies as opposed to the ones obtained from the DMFT lattice susceptibility.} In contrast, at $T=0.5$ the DF corrections overshoot the benchmark, leading to a clear deviation of about 5\%. 
This example also shows that the inclusion of non-local correlations may, but need not, result in a commensurate improvement of local correlation functions.

\subsection{Doped Hubbard model} \label{sec:doped}

\begin{figure}[t]
    \centering
    \includegraphics[width=.95\columnwidth]{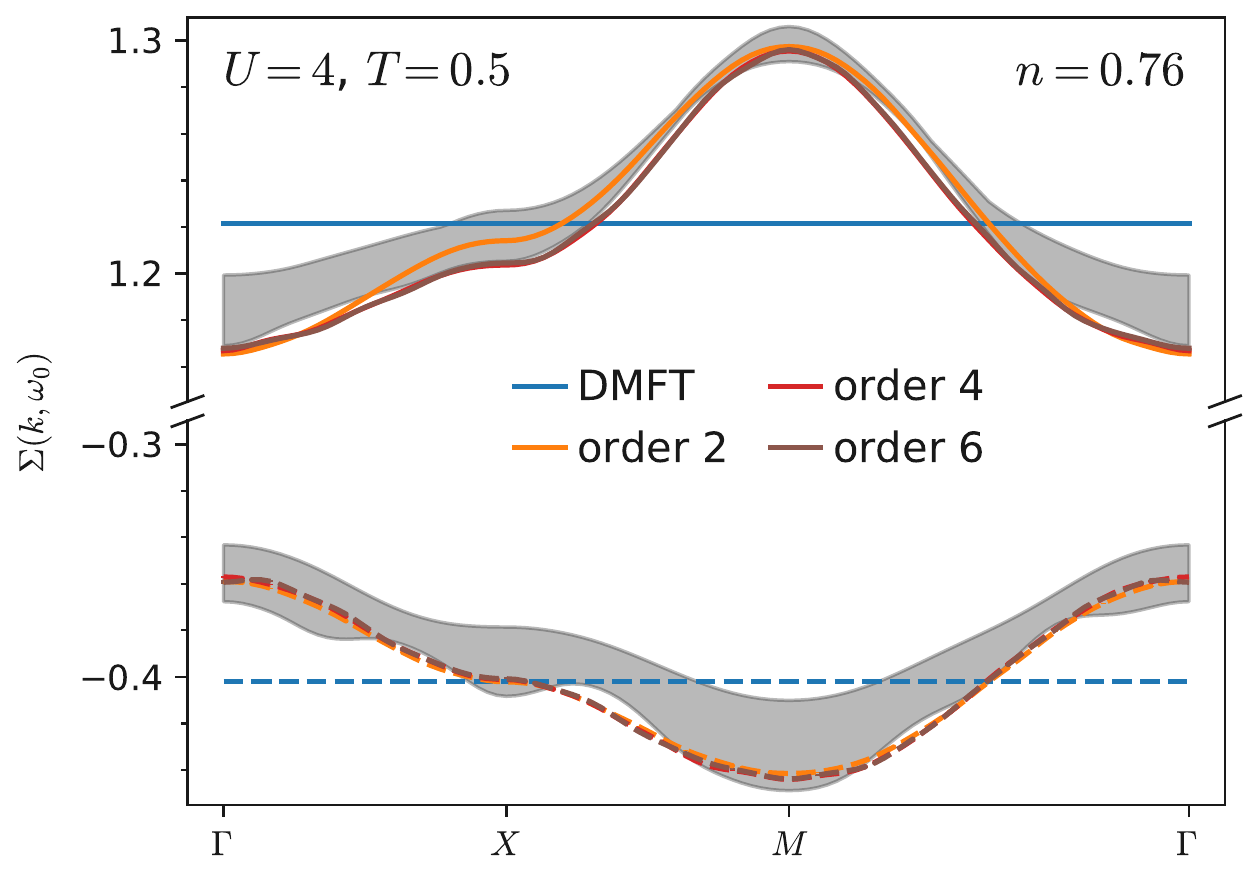} 
    \caption{Lattice self-energy at 24\% doping for $U=4$ and $T=0.5$. 
        Grey bands indicate the estimate from conventional (weak-coupling) DiagMC calculations. Different colors represent DiagMC@DF data with varying order cutoff. Real and imaginary parts are indicated by solid and dashed lines, respectively. Stochastic errors are displayed as vertical bars, but often indiscernible because they are smaller than the line width.}
    \label{fig:u4_n0.76-sigma}
\end{figure}

\begin{figure}[ht]
    \centering
    \includegraphics[width=.95\columnwidth]{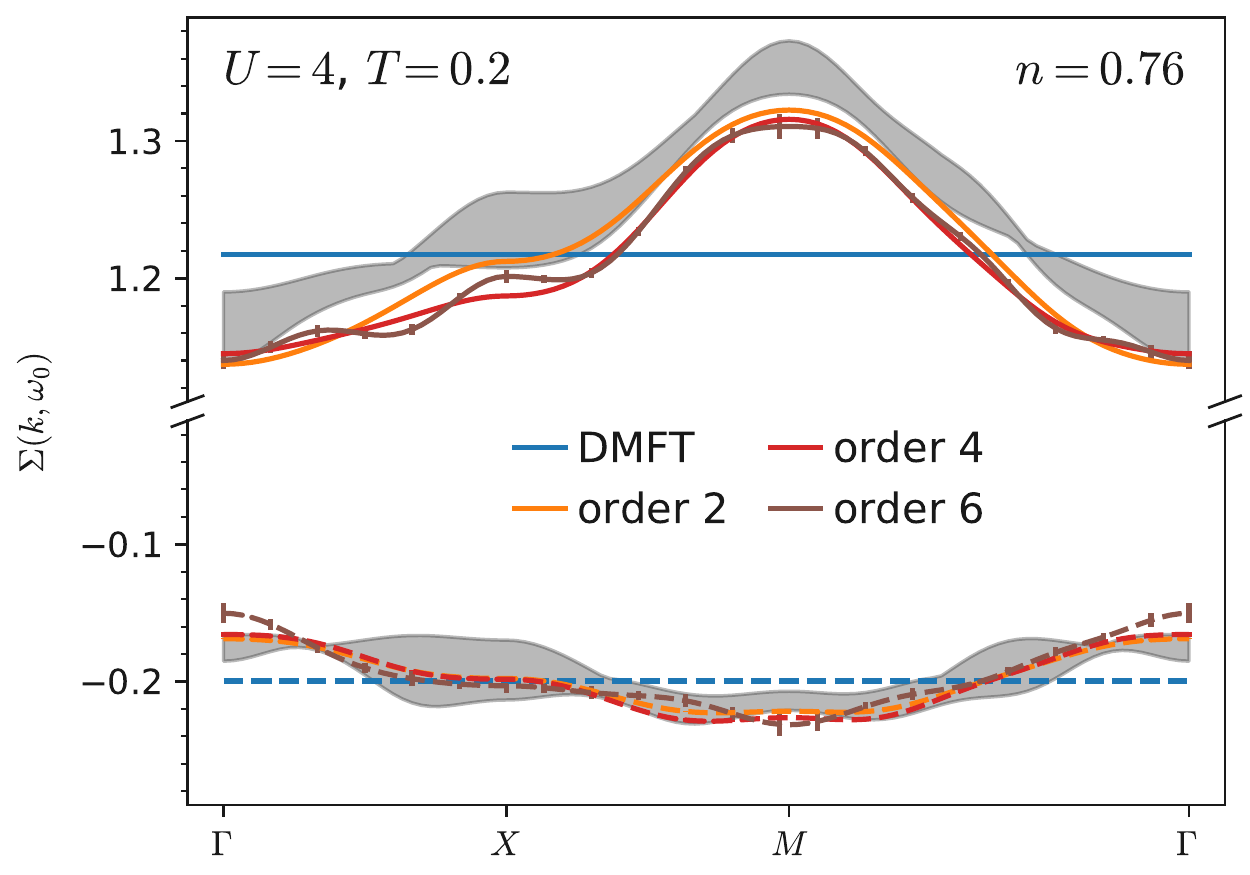}
    \includegraphics[width=\columnwidth]{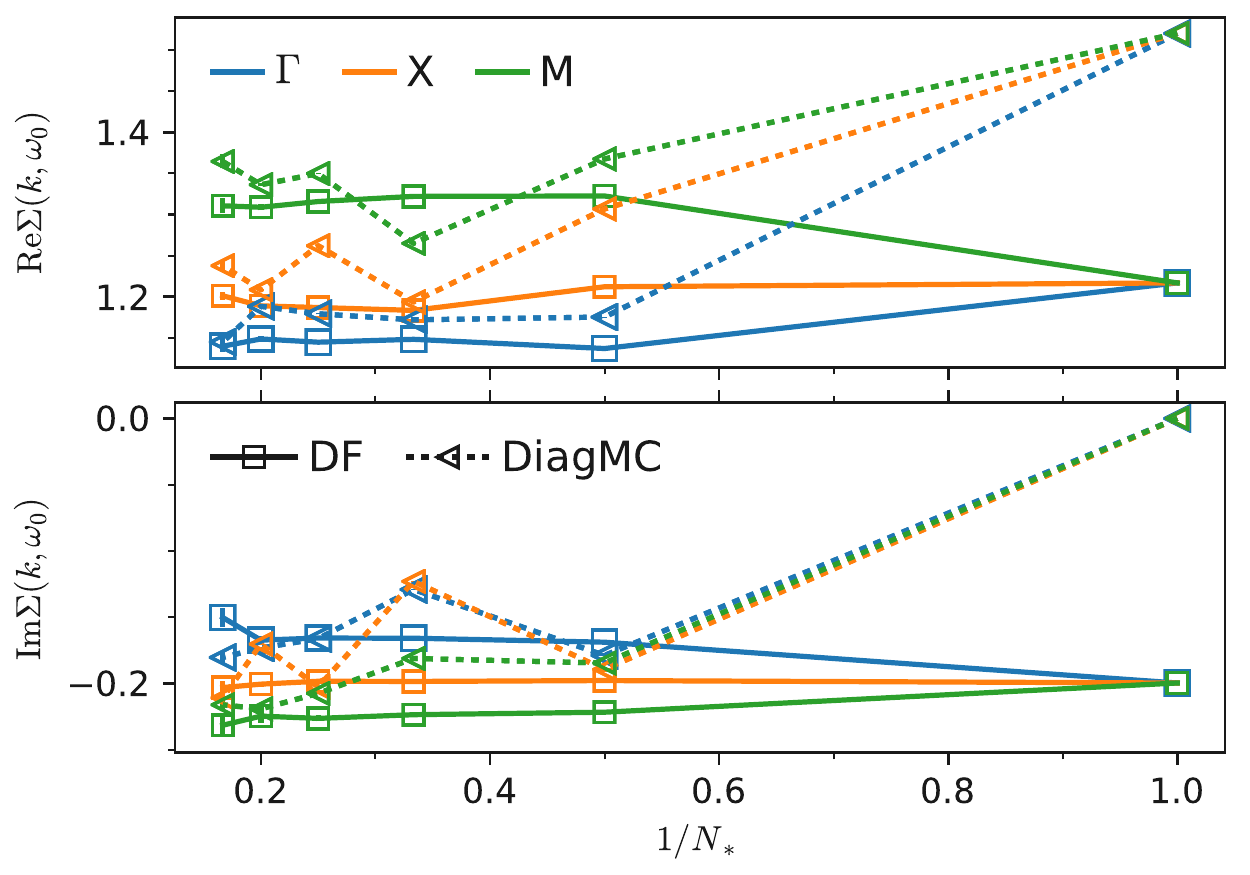}
    \caption{Lattice self-energy at 24\% doping for $U=4$, $T=0.2$. 
        Top: Momentum dependence of $\Sigma(\omega_0)$, see Fig.~\ref{fig:u4_n0.76-sigma} for an explanation of the different lines. 
        Bottom: Convergence at the high-symmetry points $\Gamma$, X, M (blue, orange, green) with diagram order $N_*$ in the DF series (squares connected by solid lines) and in the weak-coupling expansion sampled by conventional  DiagMC (triangles, dashed lines).}
    \label{fig:u4_doped-sigma_convergence}
\end{figure}

The DiagMC@DF method can be applied without modification to systems away from half filling. 
The only caveat is that non-local corrections generally give rise to corrections to the charge density, so, if a fixed density is to be targeted, the chemical potential needs to be adjusted self-consistently.
Since DDMC calculations are not feasible for the doped system due to the sign problem, we first concentrate on the case of moderate interaction $U=4$, where unbiased benchmarks can be obtained using the standard DiagMC technique.~\cite{VanHoucke2010, kozik2010} It is based on the diagrammatic series in the bare coupling $U$ and non-interacting Green's function $G_0$ (weak-coupling expansion), which converges in the regime of weak to intermediate interactions. 
Results for a system at density $n=0.76$ are presented in Figs.~\ref{fig:u4_n0.76-sigma} and \ref{fig:u4_doped-sigma_convergence} for two different temperatures.
In the latter case we also plot the DiagMC@DF and conventional DiagMC results for a few selected momenta versus diagram order cutoff $N_*$. The DF expansion starts from the momentum-independent but frequency-dependent DMFT self-energy. In the weak-coupling expansion, on the other hand, the contribution at first order is the momentum- and frequency-independent Hartree term $\Sigma_H = U n_\sigma$.
Apparently, the DF series converges much quicker than the weak-coupling series. 
Within the stochastic and systematic uncertainties, both the weak-coupling and DF series converge to the same solution homogeneously for all momenta.~\footnote{We note that a possible small constant offset in the real part of the self-energies may be caused by inaccuracies in the Hartree term due to small differences in the effective densities of the system. Also, small-amplitude oscillations visible in higher-order results are the effect of stochastic noise in the coefficients of our momentum-space basis, c.f.\ Sec.~\ref{sec:kbasis}.}

It is also interesting to compare convergence of the DF expansion at and away from half filling for the same interaction and temperature, e.g.\ the top panels of Figs.~\ref{fig:lowT-sigma} and \ref{fig:u4_doped-sigma_convergence}: At half filling, non-local corrections to the self-energy are at a similar scale as its local part and high diagram orders contain significant corrections. In the doped case, on the other hand, non-local corrections are much smaller (note the scales of the plots) and the second order already captures the converged result up to a few percent.
Like in the metallic region at half filling, a sign problem appears at low temperature and prevents us from investigating whether the DF series convergence slows down at even lower temperatures.

\begin{figure}[ht]
    \centering
    \includegraphics[width=\columnwidth]{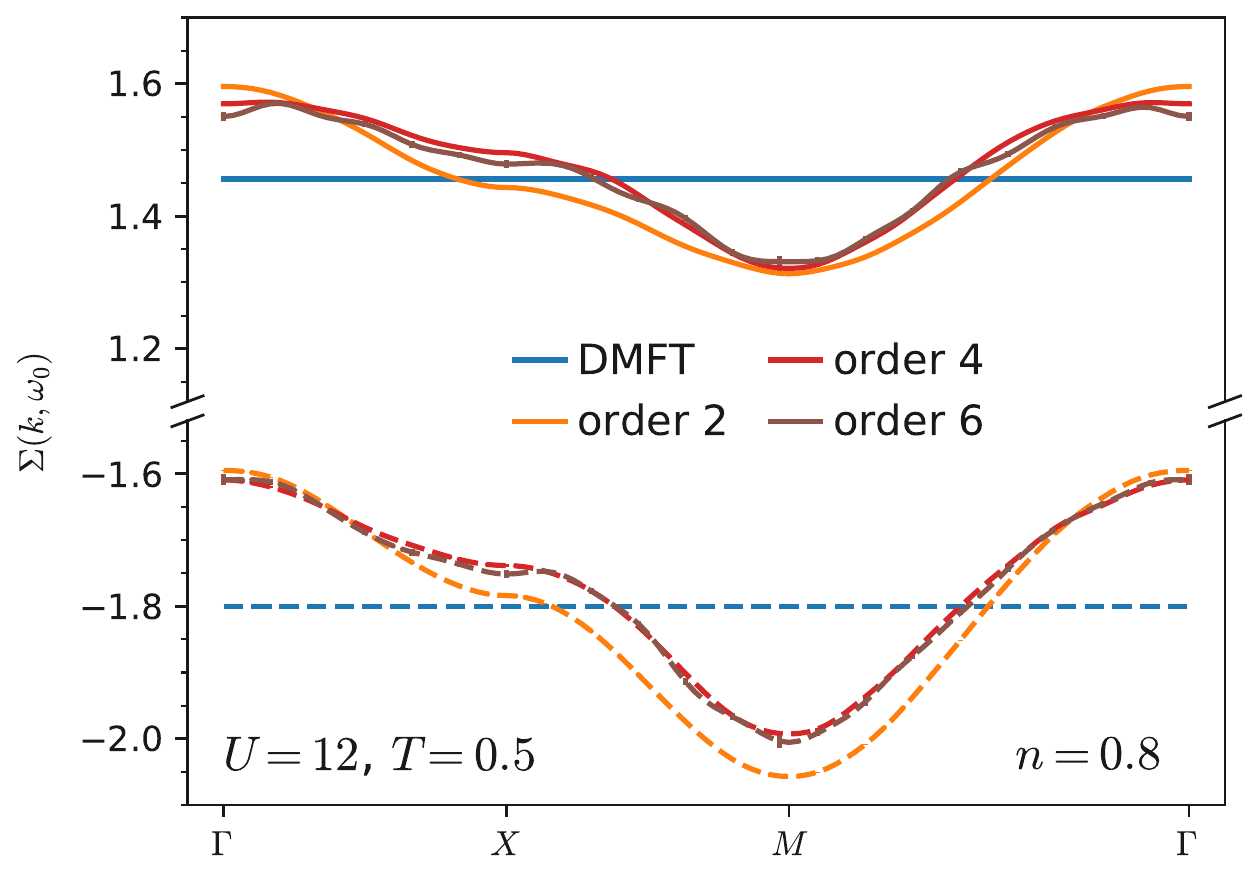}\\
    \caption{Lattice self-energy at 20\% doping for $U=12$, $T=0.5$. Different colors represent DiagMC@DF data with varying order cutoff. Real and imaginary parts are indicated by solid and dashed lines, respectively. Stochastic errors are displayed as vertical bars, but often indiscernible because they are smaller than the line width.}
    \label{fig:large_U_doped-sigma-highT}
\end{figure}

\begin{figure}[ht]
    \centering
    \includegraphics[width=\columnwidth]{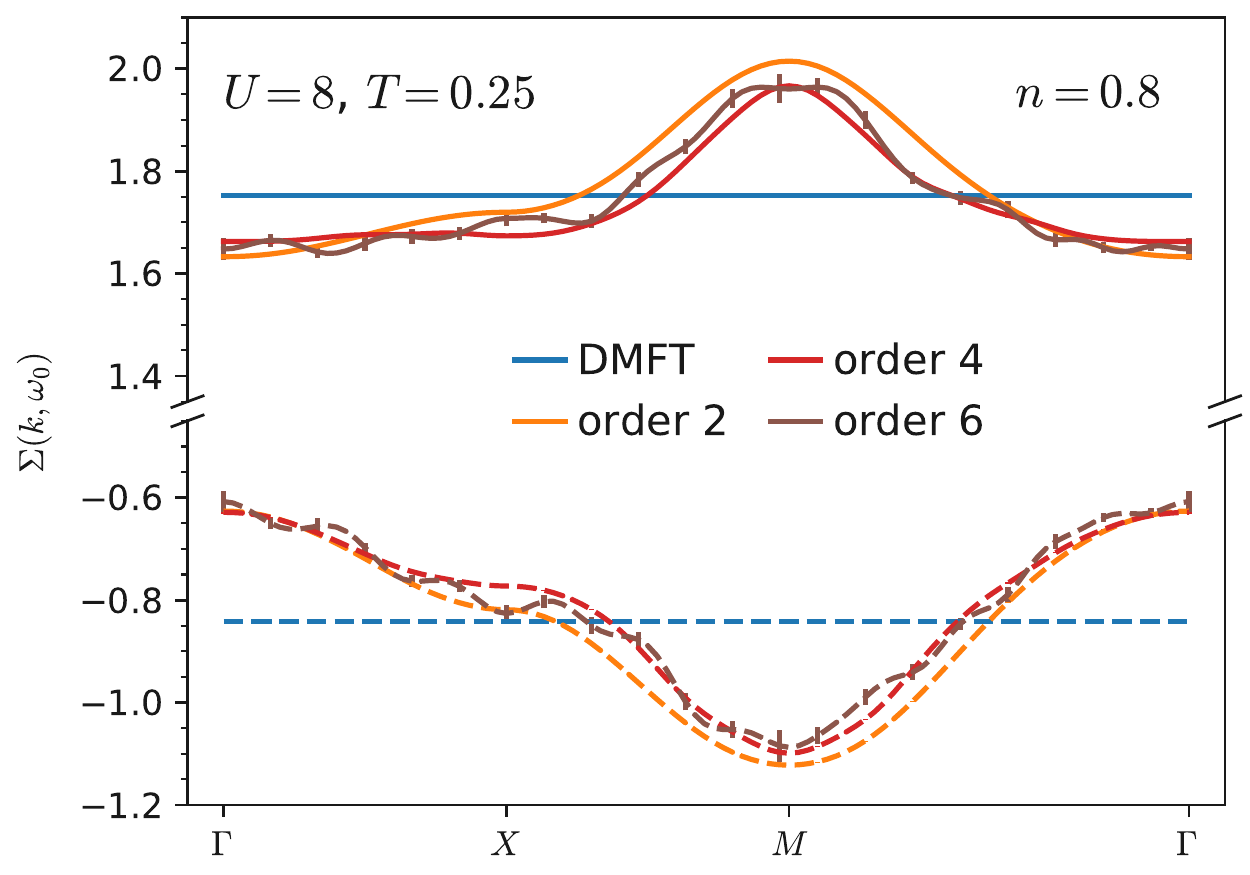}
    \includegraphics[width=\columnwidth]{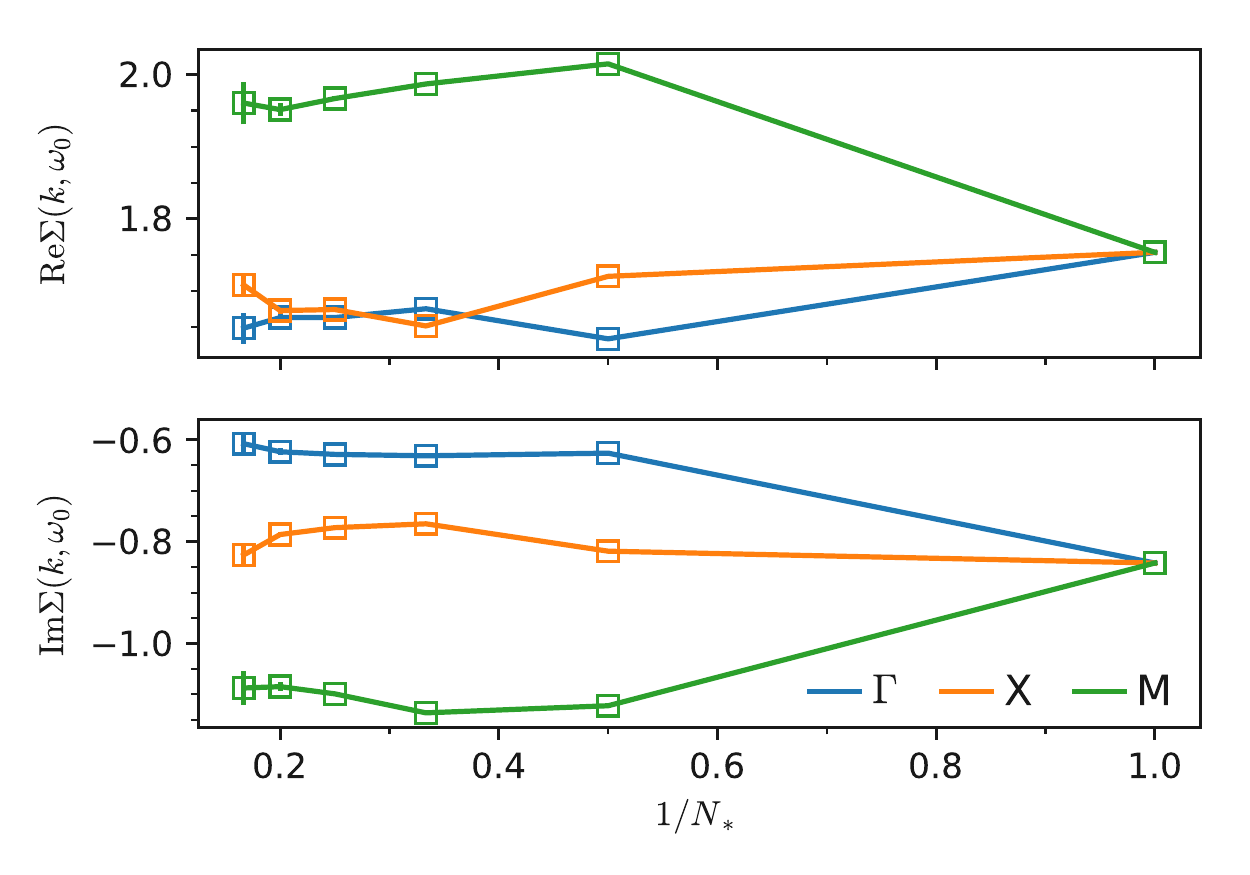}\\
    \caption{Lattice self-energy at 20\% doping for $U=8$, $T=0.25$. Top: Momentum dependence, see Fig.~\ref{fig:large_U_doped-sigma-highT} for an explanation of the different lines. Bottom: Convergence with diagram order $N_*$ at the high-symmetry points.}
    \label{fig:large_U_doped-sigma-lowT}
\end{figure}

At stronger interaction, the weak-coupling series diverges very quickly. In contrast, the DF series still features rapid convergence as demonstrated for two examples in Figs.~\ref{fig:large_U_doped-sigma-highT} and \ref{fig:large_U_doped-sigma-lowT}: In both cases the second order already contains the major momentum-dependent correction to the DMFT self-energy and orders $N_*=4$ to 6 are essentially indistinguishable within the stochastic errors. (We note that at lower temperatures, like the example of Fig.~\ref{fig:large_U_doped-sigma-lowT}, the sampling noise in the highest order is again rather large.) 
Similar to the weak-coupling case discussed above, the comparison to the half-filled case shows that doping significantly improves the convergence properties of the DF series at low temperatures: At $T=0.25$, the series for the half-filled model diverges or converges only slowly for $U=8, 12$, whereas for a doping of 20\% the series converges essentially as quickly as at $T=0.5$ for both values of $U$ (only $U=8$ is shown). At $U=12$ we find rapid convergence already at a moderate doping of 12.5\% (not shown). These findings corroborate the hypothesis that series divergence at half filling is caused by the development of strong magnetic correlations, which are effectively suppressed by doping.

\begin{figure}[htb]
    \centering
    \includegraphics[width=\columnwidth]{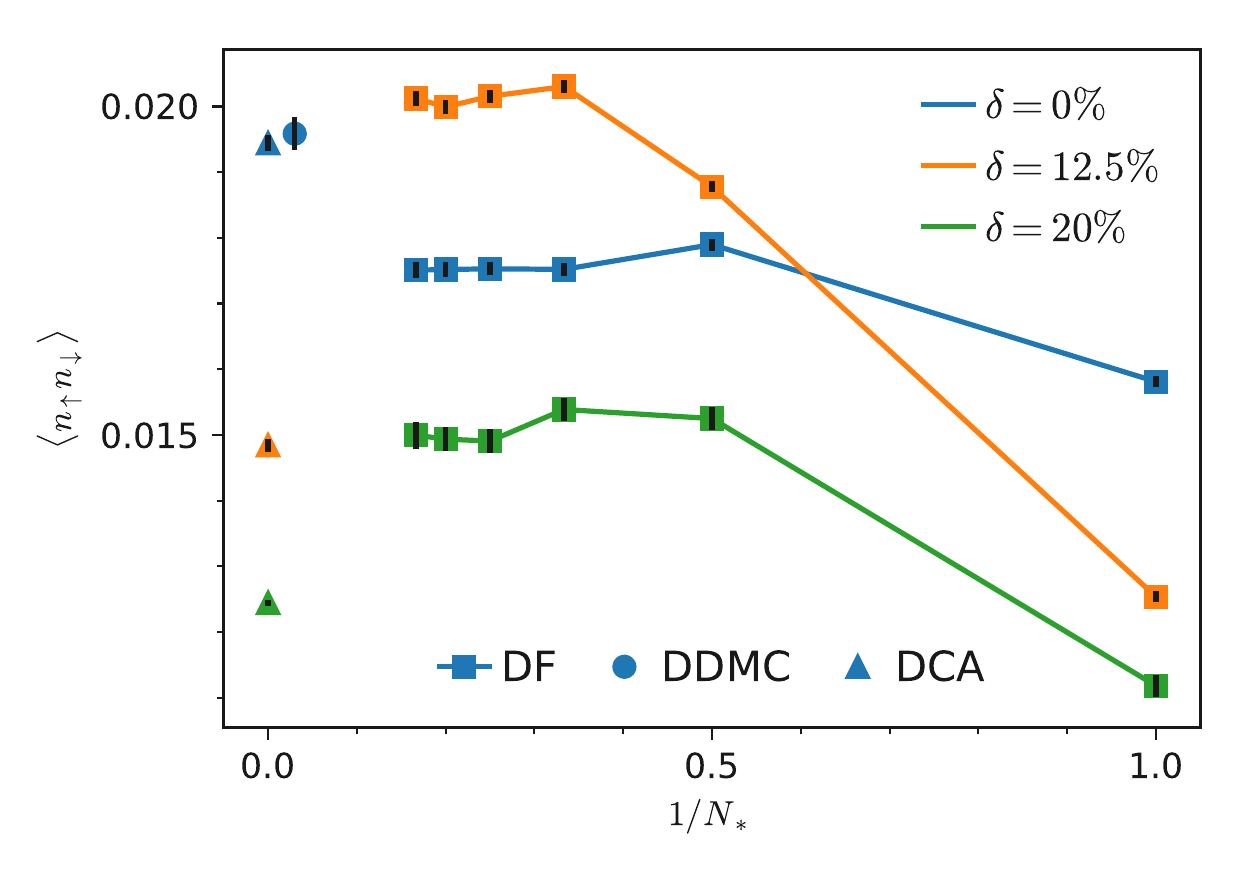}
    \caption{Double occupancy versus order cutoff $N_*$ at $U=12$, $T=0.5$ for three different values of the doping $\delta = 1-n$. Points at $N_*=1$ correspond to DMFT. Triangles indicate cluster-size extrapolated DCA results from Ref.~\onlinecite{Leblanc2015}, the circle our DDMC benchmark calculation at half filling. Error estimates are indicated by black vertical lines.}
    \label{fig:docc_vs_doping}
\end{figure}

\subsubsection{Scalar observables and high-frequency tails} \label{sec:highfreq}

While doping improves series convergence at low temperatures, a close look at scalar observables such as the double occupancy draws attention to a subtlety concerning the high-frequency asymptotics of the self-energy that has not been discussed in the DF literature so far.
Figure~\ref{fig:docc_vs_doping} shows convergence of the double-occupancy with diagram order at strong coupling and moderately high temperature for three different doping levels. These can be compared to published results obtained by extrapolating dynamical cluster approximation (DCA) calculations in cluster size~\cite{Leblanc2015} and, at half filling, to our own DDMC data.
Away from half filling the DF corrections to the DMFT result are larger and strongly overshoot the respective reference values, in particular around optimal doping. This overshooting even results in a pronounced non-monotonicity in the doping dependence of the DF observable, which is absent in the reference results.

The strong increase in the DF corrections to DMFT away from half filling appears to be linked to the fact that in the absence of particle-hole symmetry the DF density, which we iterate to the target density, differs from the impurity density, which leads to a small but significant inconsistency in the high-frequency asymptotics of the DF self-energy.
It is well known that the asymptotic behavior of the Hubbard model's self-energy at large Matsubara frequencies is exactly determined by the spin density $\langle n_\sigma \rangle$:
\begin{align} \label{eq:sigma-asymp}
\Sigma_{\vec{k} \nu \sigma} &= U \langle n_{-\sigma} \rangle + \frac{U \langle n_{-\sigma} \rangle (1-\langle n_{-\sigma} \rangle)}{i \nu} + \mathcal{O}(1/(i \nu)^2) .
\end{align}
In the DF approach this tail is produced by the impurity solver. When the impurity density equals the target density, like in the particle-hole symmetric case, a quickly decaying dual self-energy $\tilde{\Sigma} \sim 1/(i \nu)^2$ naturally preserves the correct asymptotic behavior. This is in contrast to the \dga{} and 1PI approaches, where practical approximations such as the ladder approximation typically lead to incorrect high-frequency tails unless the correct asymptotics is enforced by so-called $\lambda$-corrections.~\cite{Katanin2009,Rohringer2013}
In the absence of particle-hole symmetry the dual self-energy corrections generally give rise to a correction of the density with respect to its impurity value
\begin{align}
n - n^{\text{DMFT}} &= \Tr \Re [G - g] .
\end{align}
However, we find that this density correction arises almost exclusively from the dual self-energy's values at the lowest few Matsubara frequencies. At large $\nu$, $\tilde{\Sigma}$ still decays quickly and leaves the impurity self-energy essentially unchanged.~\footnote{
    For the cases we checked, possible differences in the leading coefficients of the tail are at least an order of magnitude smaller than the density corrections would imply per Eq.~\eqref{eq:sigma-asymp}.
}

For self-consistent calculations it is in fact easy to show that the dual-fermion expansion truncated to two-particle interactions always produces a dual self-energy that decays at least as fast as $\mathcal{O}(1/\nu^2)$. The reasons are (1) all local dual self-energy diagrams, \ie diagrams where the external legs are attached to the same interaction vertex, vanish due to the self-consistency condition $\sum_{\vec{k}} \tilde{G}_{\vec{k} \nu \sigma} = 0$ and (2) all other dual self-energy diagrams cannot have a constant or $\mathcal{O}(1/\nu)$ tail due to the fast decay of the dual propagator $\tilde{G}_{\vec{k} \nu \sigma} = \mathcal{O}(1/\nu^2)$.~\footnote{
    This is most easily seen in imaginary-time space, where a constant term in the high-frequency tail corresponds to a $\delta(\tau)$ term and a $1/\nu$ term to a jump in imaginary time. Then continuity of the propagators crossing an arbitrary cut between the vertex with the incoming leg and the one with the outgoing leg immediately implies continuity of the whole diagram with respect to variations of the time difference between the in- and outgoing legs.
} 
When at least three-particle interactions are included in the expansion, there are local diagrams that do not vanish and can change the self-energy's tail, in accordance with the expectation that the DF formalism is exact independent of whether the impurity and target densities agree or not. 
In our non-self-consistent DiagMC@DF calculations, Hartree- and Fock-like diagrams with self-energy insertions on the propagator line may in principle produce corrections to the tail, too. Our numerical finding that these are small indicates that, independent of whether self-consistency is employed, high-frequency corrections to the self-energy are mostly contained in the neglected higher-order vertices.
This does not contradict our earlier finding that the effect of higher-order vertices is relatively small: Even in the 12.5\% doping case shown in Fig.~\ref{fig:docc_vs_doping}, the relative density difference -- and hence the expected correction to the high-frequency coefficients -- is only $(n - n^{\text{DMFT}})/n = 2.5\%$.
But considering the importance of the self-energy's high-frequency behavior for sum rules and the evaluation of scalar observables a better treatment would clearly be desirable.

\subsection{Diagram topologies} \label{sec:topologies}

The DiagMC@DF technique can provide information about which diagram topologies prevail in the sampling process: Since the frequency of a specific topology in the Markov chain is directly proportional to its weight (i.e.\ the absolute value of the integrand, averaged over internal variables), configurations with a large impact on the result should be sampled frequently (importance sampling). 
This may in principle provide guidelines, e.g., for the construction of computationally cheaper approximations. 
We note that the choice of updates does not influence the weight or frequency of a configuration and hence the topology of the sampled diagrams as long as it satisfies detailed balance and ergodicity.
For the analysis one needs to bear in mind, however, that, due to the sign alternation between diagrams, a large sampling frequency does not necessarily imply a large impact on the results. A good example we found is diagrams with local dual propagators which would be sampled over 90\% of the time (if not suppressed, c.f.\ Sec.~\ref{sec:optim}) although their integral is zero.

In spite of this caveat we examined the topologies for one specific set of parameters: $U=4$, $n=1$ and $T=0.5$. We find that the relative frequency of sampling a topology decreases quickly (presumably exponentially) with diagram order. Since the number of topologies increases factorially with diagram order, the sampling process spends overall more time at the higher orders. Among the higher orders (four or above), the most frequent topologies are Hartree and Fock diagrams where the propagator is dressed with complicated self-energy insertions such as particle-hole ladders and bubbles. Their frequencies are an order of magnitude larger than those of skeleton diagrams of the same order. If one only considers skeleton diagrams, the most frequent topology at fourth and fifth order is the particle-particle ladder. However many other, more complicated, topologies have only slightly lower frequencies. Due to the large number of possible topologies, generic topologies are sampled most of the time and no particular topology stands out. In particular we do not find that ladder diagrams prevail.

\subsection{Role of one-particle reducible diagrams and denominator in the self-energy formula} \label{sec:denominator}

\begin{figure}[tpb!]
    \centering
    \includegraphics[width=\columnwidth]{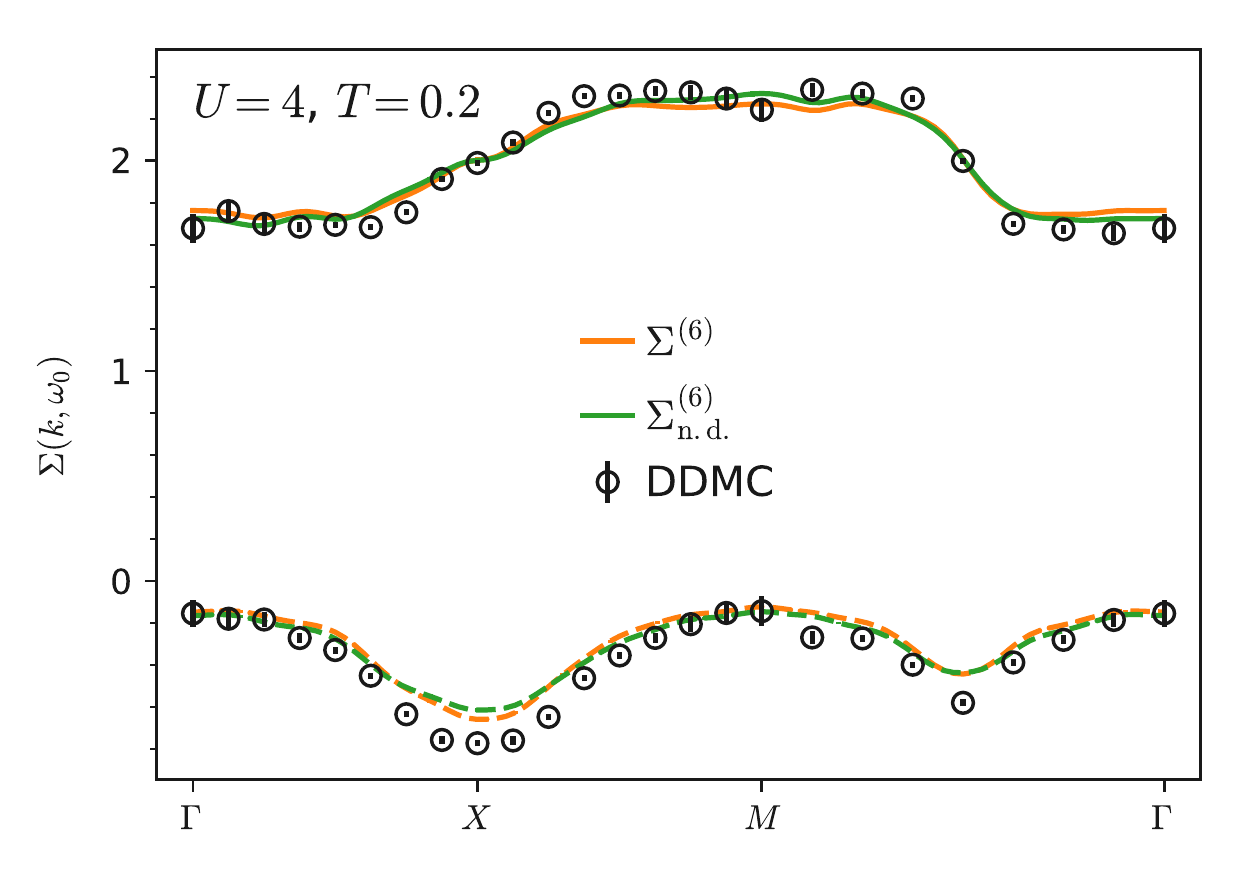}
    \includegraphics[width=\columnwidth]{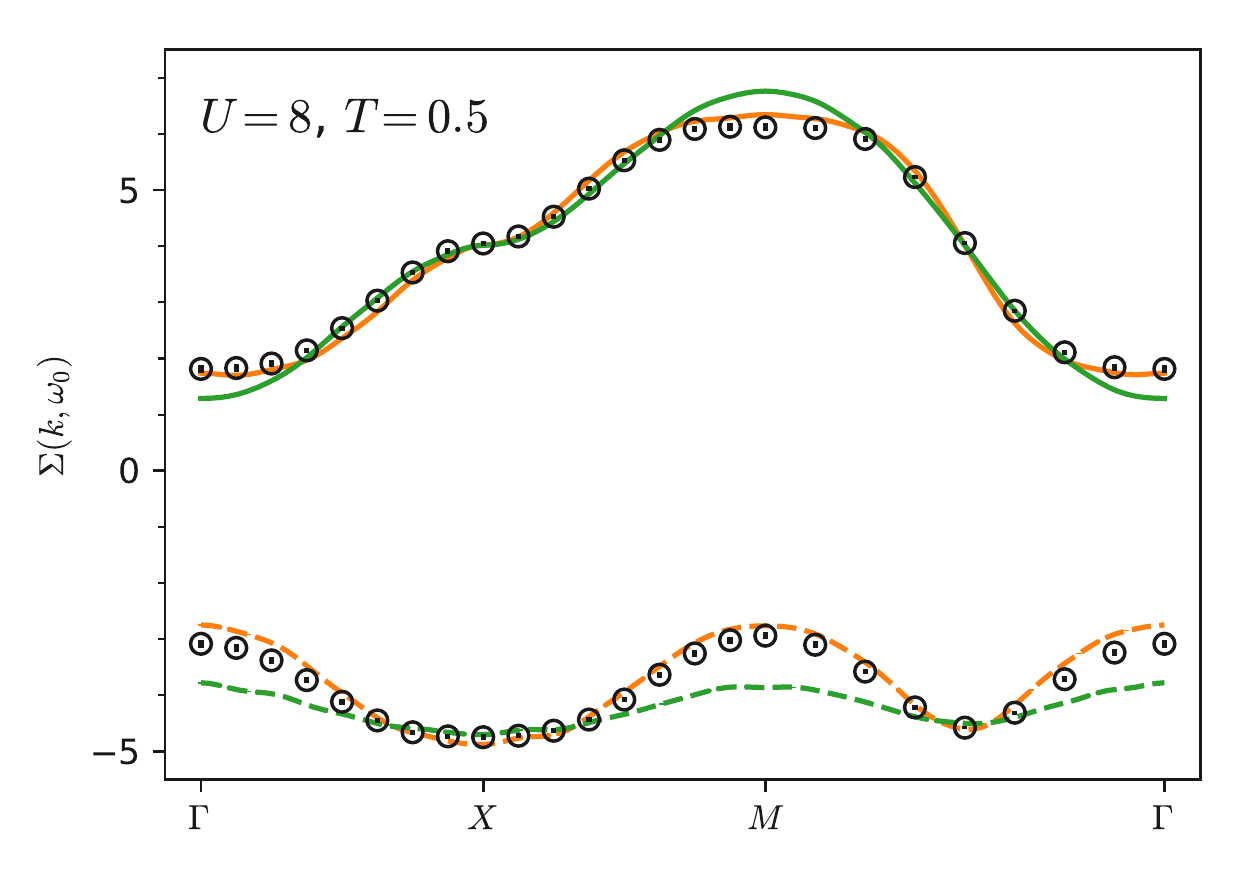}
    \includegraphics[width=\columnwidth]{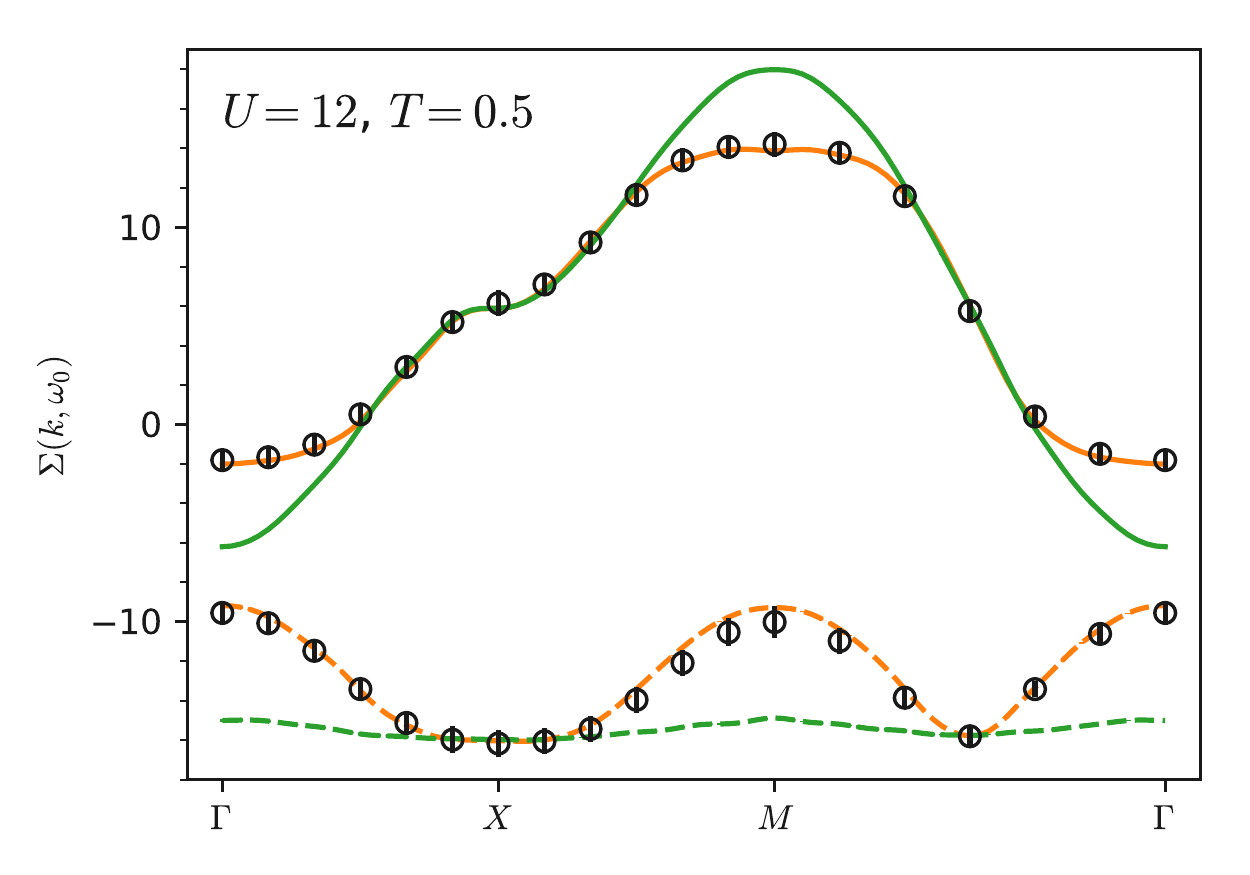}
    \caption{Modified lattice self-energy (green lines) calculated according to Eq.~\eqref{eq:sigma-nodenom} compared to the original self-energy (Eq.~\eqref{eq:sigma-from-sigmad}, orange lines) and DDMC benchmarks for three sets of parameters at half filling. In all cases the cutoff order is $N_*=6$. Self-energy data from different orders and ladder approximations for the same parameters can be found in Fig.~\ref{fig:lowT-sigma} (top) and Fig.~\ref{fig:u8n12_b2_n1-sigma}, respectively. Stochastic errors are displayed as vertical bars on all data, but often indiscernible because they are smaller than the line width.  
    }
    \label{fig:sigmatilde}
\end{figure}

While the dual-fermion expansion in principle provides an exact expression for the lattice Green's function, the validity of its restriction to two-particle interactions (DF2P approximation) has been questioned.~\cite{Katanin2013} This has motivated the development of the one-particle irreducible approach (1PI).\cite{Rohringer2013}
Numerical results indicate that including diagrams containing the three-particle vertex has a small effect over a broad range of interaction values.~\cite{Hafermann2009} The importance of higher-order interactions in general is a priori unclear. It has been argued~\cite{Katanin2013} that the denominator in Eq.~\eqref{eq:sigma-from-sigmad} relating the dual- and lattice-fermion self-energies should be used with care. The reason is that it generates terms that are reducible with respect to the impurity Green's function and that may or may not be exactly canceled by corresponding contributions contained in higher-order vertices.
Cancellation happens when the one-particle irreducible (1PI) contributions of these vertices are negligible. In this case neglecting the three-particle vertex means the denominator will not be cancelled even though it should be. If this were to happen, the denominator should be excluded from a DF2P calculation. The other possibility is that the sum of diagrams containing three-particle or higher-order vertices may be small as a whole. Then, neglecting them will have a small effect and the denominator is important. This raises the important practical question of when the denominator should be used in DF2P.

To answer this question, we compute the physical self-energy neglecting the denominator in Eq.~\eqref{eq:sigma-from-sigmad}. The resulting lattice self-energy
\begin{align} \label{eq:sigma-nodenom}
    \Sigma_{\text{n.d.}} &= \Sigma_{\text{imp}} + \tilde{\Sigma}
\end{align}
is presented in Fig.~\ref{fig:sigmatilde} for three examples alongside the original self-energy calculated with the same order cutoff. At strong coupling the result compares significantly worse with our exact benchmarks than the original one obtained from Eq.~\eqref{eq:sigma-from-sigmad}, which supports the conjecture and previous evidence~\cite{Hafermann2009} that contributions to $\tilde{\Sigma}$ containing higher-order vertices are indeed small in this regime.
At weak coupling the two alternative self-energies can hardly be distinguished. For $U=4$, the denominator is always very close to 1. The largest deviations from unity at $T=0.2$ are ~2\% and only ~0.3\% at $T=0.5$. By contrast, at $U=8$ and $U=12$ the deviations reach 30\% and 70\%, respectively, already at $T=0.5$.
Considering the case of $U=4$, $T=0.2$, shown in the top panel of Fig.~\ref{fig:sigmatilde}, the drop of the denominator appears to deteriorate rather than improve the convergence of $\Im \Sigma$ towards the benchmark data. The effect is small and neither case is fully converged with diagram order, so that our results do not conclusively rule out either possibility.
Nevertheless, we have not seen a single case so far where dropping the denominator visibly improves the result. It should be included in DF2P calculations by default.

Ref.~\onlinecite{Rohringer2013} includes a comparison of results produced by ladder-approximations to the DF, 1PI, and \dga{} approaches. Some significant differences between the methods are apparent, but due to the lack of reference data the different approximations' respective qualities have remained unclear so far.
In this work we have established a reference by performing numerically exact DDMC calculations. We have found good agreement between converged DF2P results and the DDMC benchmarks for similar parameters \footnote{
    Rohringer \etal compare one-shot and self-consistent ladder calculations of 1PI, DF and \dga{} at $U=4$, $T=0.235$ and $U=8$, $T=0.444$ in units of $t$. The closest points where we have benchmarked the dual-fermion approach are $U=4$, $T=0.2, 0.25$ and $U=8$, $T=0.4, 0.5$, respectively. Note that the unit of energy in Ref.~\onlinecite{Rohringer2013} differs by a factor of four from our convention.
} as those used in Ref.~\onlinecite{Rohringer2013}.
One is hence drawn to the conclusion that, even though additional diagrams are included in 1PI, the quality of the approximation may be worse in some relevant cases. 
It is clear that the inclusion of more diagrams in general does not necessarily improve the quality of a many-body theory and a more systematic comparison of the performance of different diagrammatic extensions of DMFT is highly desirable. DiagMC sampling allows one to perform such a comparison without biasing the results by a restriction to specific diagram topologies.

\begin{figure}[ht]
    \centering
    \includegraphics[width=\columnwidth]{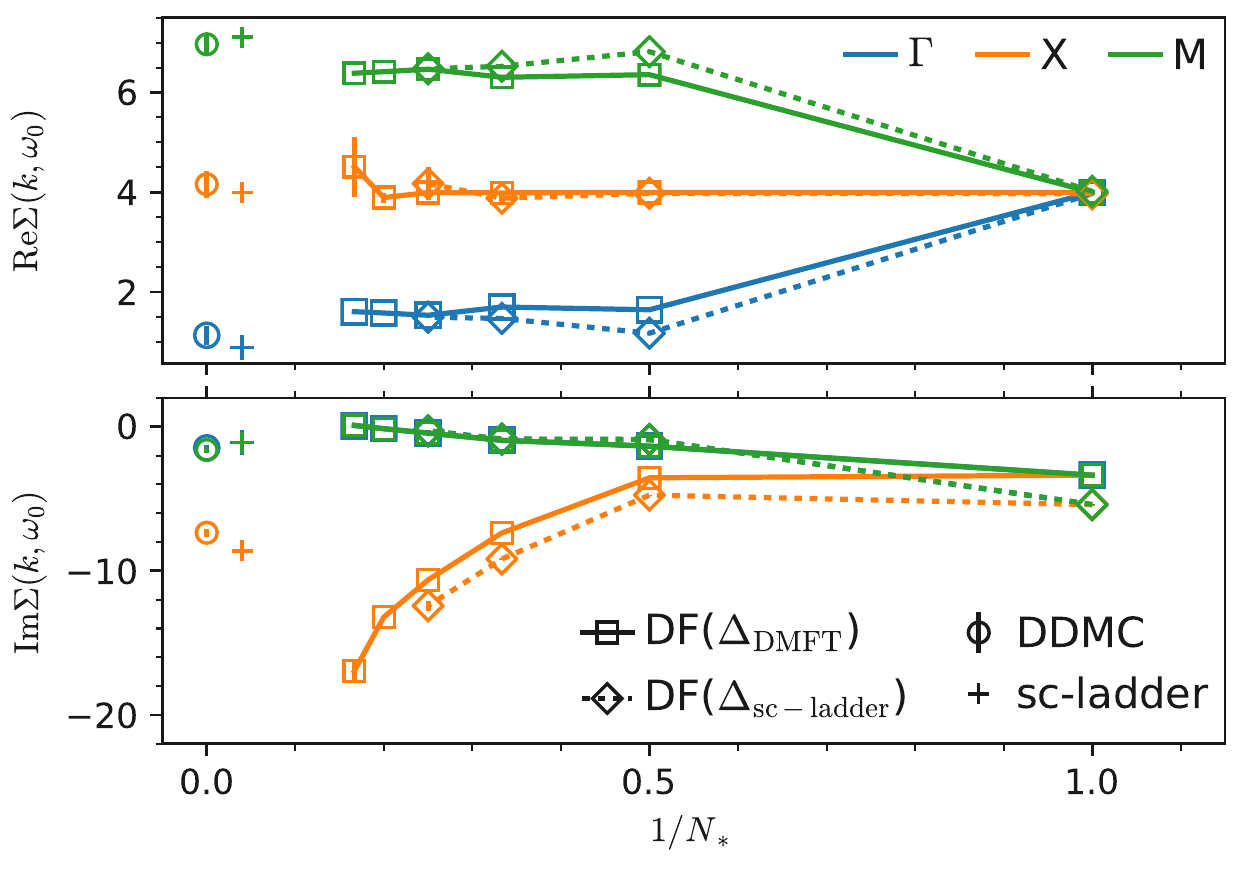} 
    \includegraphics[width=\columnwidth]{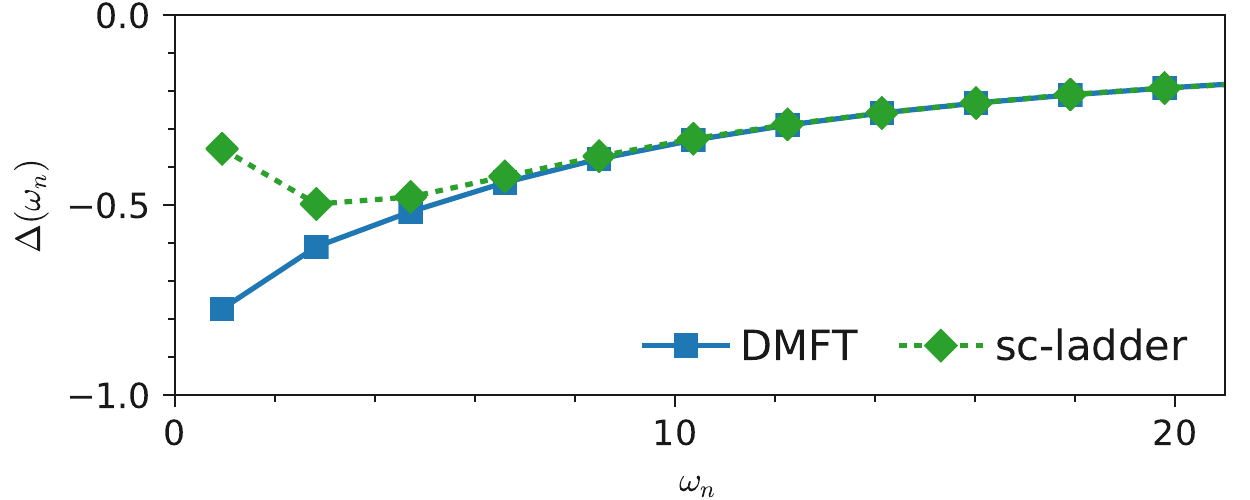}
    \caption{Top and center: Lattice self-energy computed from different hybridization functions for the half-filled system at $U=8$ and $T=0.3$. Shown is the dependence on cutoff order $N_*$ at the high-symmetry points $\Gamma$ (blue), X (orange), and M (green). Points connected by solid (dashed) lines correspond to an expansion around DMFT (self-consistent DF ladder approximation). Both show a divergence in the imaginary part at the X point. For comparison, the corresponding DDMC and sc-ladder results are indicated, too. Bottom: DMFT and sc-ladder hybridization functions that have been used as inputs for the two different DiagMC@DF calculations.}
    \label{fig:u8_b3.333_scl-sigma}
\end{figure}

\subsection{Choice of hybridization function}
\label{sec:hybchoice}

In the DF formalism, the hybridization function $\Delta(i \omega_n)$ appears as a free parameter, which can in principle be tuned to improve the series convergence behavior.
In this case, the expansion is no longer around the DMFT solution, but around an auxiliary -- and possibly more optimal -- reference impurity problem.
This is routinely done in self-consistent ladder DF calculations, for instance, and one may hope that also for DiagMC@DF the observed convergence problems at low temperatures might at least be ameliorated by an optimized choice of $\Delta$.
While the input of a different hybridization function in itself does not require any changes to the sampling code, there is still a technical complication. Namely if the hybridization is not the DMFT one, the bare DF propagator has a local part $\int dk \tilde{G}^{(0)}(\vec{k}, i \omega_n) \neq 0$, so that diagrams with local propagators cannot be discarded any more.
The corresponding sampling of additional diagram topologies turns out to considerably increase the computational effort needed to converge the stochastic errors.
A more fundamental issue, however, is the fact that we have found no example where the series convergence behavior could be significantly improved by tuning the hybridization function.
The case at $U=8$, $T=0.3$ shown in Fig.~\ref{fig:u8_b3.333_scl-sigma} exemplifies this: here the DMFT hybridization function is metallic whereas the hybridization function resulting from a self-consistent ladder DF calculation shows a clear insulating downturn at low frequencies, but the order-by-order results for the self-energy are very similar for the two corresponding expansions.
This indicates that the breakdown of the DF series, at least at half filling, is not primarily caused by an incorrect metallic or insulating character of the starting point, but likely due to the buildup of strong magnetic correlations (see our discussion in Sec.~\ref{sec:tbreakdown}).
However, we can not rule out that the divergence is of a purely mathematical nature unrelated to any physics, which is a typical case for the bare series of the original Hubbard model, as discussed in Sec.~\ref{sec:overview}.

An alternative way of taking advantage of the freedom of choice in the hybridization function arises in the absence of particle-hole symmetry, where differences in the charge density between impurity and dual-fermion results lead to issues with the high-frequency asymptotics of the self-energy and the evaluation of scalar observables as described above in Sec.~\ref{sec:highfreq}. Here a suitable choice of hybridization function might alleviate the problem by reducing, or even eliminating, the density difference.

On a related note, another possibility to improve the DiagMC@DF scheme similar to the change of hybridization function is dressing the dual propagator with a non-local self-energy and iteration of the sampling process to self-consistency, similar to, e.g., a fully self-consistent ladder calculation.
Technically, such a change is straight-forward. One only needs to restrict the sampling process to skeleton diagrams, which can be achieved by dropping diagrams with self-energy insertions at the time of measurement; these can be efficiently detected by checking for propagators with identical momenta, c.f.\ Ref.~\onlinecite{gukelberger2015diss}. 
In this work we have restricted ourselves to one-shot calculations and it is not clear whether this type of self-consistency would improve the convergence behavior of the expansion. 
In the standard diagrammatic technique, self-consistency does not always bring improvement of convergence properties. Skeleton expansion in terms of the full Green's function $G$ often exhibits slower convergence than the expansion in terms of the bare Green's function $G_0$ or can even converge to unphysical solutions.~\cite{kozik2015nlf}

\section{Discussion} \label{sec:discussion}

We have presented an approach that combines a diagrammatic extension of DMFT with diagrammatic Monte Carlo sampling in order to go beyond the leading-order or ladder approximations that are commonly used in practical implementations of these extensions. Specifically, we have laid out in detail a DiagMC algorithm that computes the dual-fermion expansion to high orders, including arbitrary diagram topologies involving two-particle interaction vertices. Crucially, this allows one to judge whether and how well the DF series converges in different parameter regimes and provides an estimate of the systematic error introduced by the truncation of the series to low orders or specific classes of diagrams.
Thus we could systematically investigate the convergence properties of the DF series for the self-energy in different parameter regimes. 


Our extensive comparisons to numerically exact benchmarks show that, from high temperatures down to the vicinity of the DMFT Néel transition, the dual-fermion series converges very quickly to the exact solution over the full interaction range. Upon lowering the temperature further, however, we generically observe slower series convergence or convergence towards an incorrect (though often qualitatively reasonable) solution, and ultimately series divergence. The studied examples in the vicinity of half filling provide evidence that the buildup of substantial magnetic correlations, which cannot be captured by single-site DMFT, leads to a breakdown of the DF expansion at low temperatures.

In those regimes where the self-energy was observed to converge quickly with expansion order, a remarkable agreement with the exact benchmarks shows that higher-order interaction vertices do not play a significant role.
Here also the popular self-consistent ladder approximation yields highly accurate results, and in the crossover regime where the series starts to break down, the self-consistent ladder often yields at least qualitatively reasonable results for the self-energy.
We suggest that, in cases where DiagMC@DF results are not available, the difference between one-shot second-order and self-consistent ladder calculations can give a reasonable estimate of the quantitative error in the self-energy.
In particular away from half filling we further found that scalar observables may exhibit significant deviations from our benchmarks even in regimes where the self-energy is still quite accurate.
Despite the fact that the overall effect of neglecting higher-order vertices on the self-energy is small in many regimes, it is desirable to improve the evaluation of scalar observables when these are neglected.

\subsection{Outlook}

For the low-temperature regime where DiagMC@DF results are not satisfactory, possible extensions may be considered, namely the inclusion of higher-order vertices, an improvement of the reference solution, or sampling of a series in terms of dressed propagators. The first is straightforward in principle, and the question whether the convergence problems we found can be linked to truncation of the dual-fermion action certainly deserves further studies. However, systematic inclusion of $n$-particle vertices with $n=3, 4, \dots$ would be computationally challenging due to the exponential growth of the computational cost of handling higher-order vertices, and the inclusion of additional interactions would not cure the basic problem that a single impurity atom cannot represent non-local correlations and thence might not be the best starting point for a perturbative expansion when such correlations are large.
Therefore a more promising route for practical calculations may be to base the DF expansion on an improved reference solution, e.g.\ by considering cluster impurities, which can capture at least short-range magnetic correlations already on the impurity level.~\cite{Yang2011a}  Finally, the sampling in terms of dressed propagators might remove or at least alleviate the breakdown of the series.

Extension of the DiagMC@DF method to sample other diagrammatic quantities such as two-particle vertex functions, which give access to susceptibilities and hence to characteristics of continuous phase transitions, is straightforward along the lines of similar implementations for the standard diagrammatic technique.~\cite{VanHoucke2012,deng2014ebr,gukelberger2015diss} 
Last but not least, the application of DiagMC sampling to other diagrammatic extensions, such as DiagMC@\dga{} and DiagMC@1PI, would allow for a systematic comparison of the convergence properties of the different diagrammatic schemes. Work in this direction is underway.
The analysis of the sampled diagram topologies in DiagMC@DF did not identify any particularly important class of diagrams. Nevertheless, for more complicated diagrammatic extensions like the dual boson (DB) approach,~\cite{Rubtsov2008} which contain two types of propagators and vertices, DiagMC@DB may provide guidelines of how to construct diagrammatic approximations.

\subsection{Related work}

\begin{figure}[tb]
    \centering
    \includegraphics[width=\columnwidth]{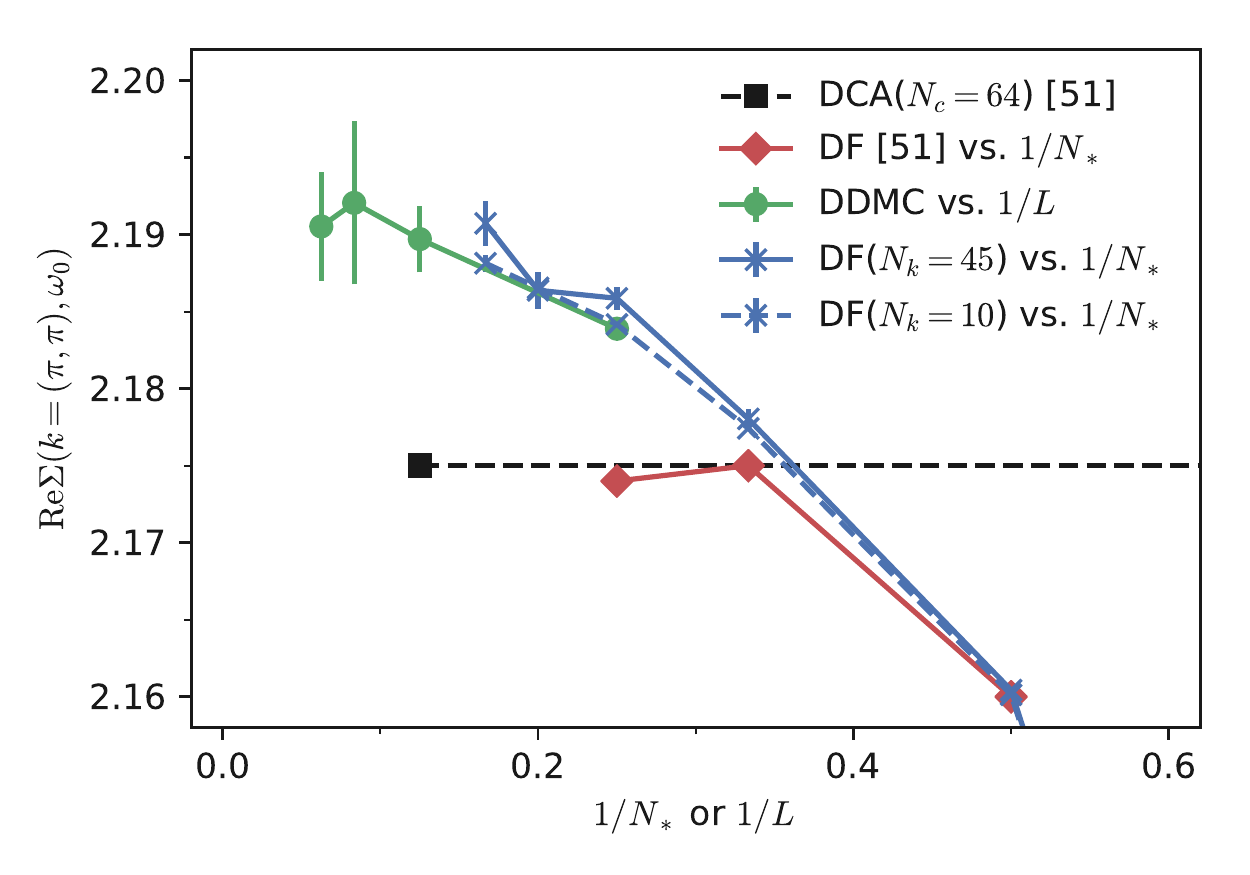}
    \caption{Lattice self-energy at the M-point. Black and red squares are DCA and DiagMC@DF data, respectively, digitized from Fig.~12 of Ref.~\onlinecite{Iskakov2016}. Blue and green symbols represent our DiagMC@DF and DDMC data. DF (DDMC) data are plotted as a function of cutoff order $N_*$ (linear system size $L$). Ref.~\onlinecite{Iskakov2016} does not provide error estimates, but their plot symbols have a radius $\sim 0.003$ on the y-axis scale. Blue crosses connected by dashed lines indicate the effect of reducing the number of momentum basis functions from $N_k=45$ to $N_k=10$.}
    \label{fig:u4_b2-sigmare-comparison}
\end{figure}

While our benchmark calculations were running, an independent work on the new method appeared.~\cite{Iskakov2016} 
Our present work additionally includes a systematic comparison to numerically exact benchmarks for different interactions and temperatures and thus allowed us to map out the method's range of applicability and draw conclusions on the relevance of higher-order vertices, the choice of hybridization function, and the issue of the self-energy denominator. 
We cannot directly compare results for the dual self-energy between the two works because they depend on the exact choice of the hybridization function. However, we can compare data for the lattice self-energy at the parameters $U=4$, $T=0.5$, and half filling as shown in Fig.~\ref{fig:u4_b2-sigmare-comparison} for one momentum $k=(\pi, \pi)$. We see that, while the overall results are very similar, there are some quantitative differences. The second- and third-order results agree within the symbol size of Ref.~\onlinecite{Iskakov2016}, whereas fourth-order results differ noticeably: In Ref.~\onlinecite{Iskakov2016} the $N_*=3$- and $N_*=4$-data are indistinguishable in the neighborhood of the $\Gamma$- and M-points, but our results show a correction of the order of 5\% to $\Re \Sigma-\mu$ when going from third to fourth order. 
In Fig.~\ref{fig:u4_b2-sigmare-comparison} we also indicate the effect of reducing the number of momentum basis functions from $N_k=45$ to $N_k=10$ (corresponding to a reduction of $n_{\text{max}}$ from 8 to 3 in the notation of Sec.~\ref{sec:kbasis}). We can safely rule out that the finite momentum basis has a large effect on our results.
Whether the discrepancy is due to statistical errors or a more fundamental difference is unclear because statistical error bars are not specified in Ref.~\onlinecite{Iskakov2016}. Different choices of hybridization function will in general also lead to differences in order-by-order results and in the effects of higher-order vertices.
We note that even the benchmark data exhibit a very similar difference: Iskakov \etal compare to a DCA self-energy obtained on a 64-site cluster, which agrees to the third-order DF result within symbol size at all plotted cluster momenta. At the M-point both their third- and fourth-order results coincide with the DCA one $\Re \Sigma(k=(\pi,\pi), \omega_0) - \mu = 0.175$ (and correspondingly for the $\Gamma$-point, which is related by particle-hole symmetry).
In our DDMC simulations, on the other hand, this value is converged to 
0.190(5) for linear system sizes $L=8, 12, 16$ and our DF results quickly approach this value with increasing order $N_*=4, 5, 6$. One likely reason for the difference between DCA and DDMC is the averaging over $k$-space patches implicit in DCA that leads to a smoothening of extrema in $\Sigma(k)$.

\section{Conclusions} \label{sec:conclusions}

We have laid out a computational method for correlated lattice systems that combines the diagrammatic Monte Carlo technique with the dual-fermion approach. Extensive benchmarks against numerically exact calculations allowed us to map out the method's range of applicability for the case of the Hubbard model on the square lattice. 
In contrast to the conventional DiagMC method, which is based on the standard weak-coupling expansion for the Hubbard model, the method is applicable over the full range of interactions from weak to strong coupling. In comparison to the popular second-order and ladder approximations to the dual-fermion approach, DiagMC sampling removes the bias inherent in the restriction to specific diagram topologies and allows one to judge the relevance of neglected higher-order terms.
Comparison of the result to benchmarks obtained from numerically exact diagrammatic determinant Monte Carlo shows that at temperatures where magnetic correlations are weak, the dual-fermion series converges very quickly to the exact solution and contributions from higher-order vertices are small. Upon lowering the temperature, however, we generically observe slower series convergence, convergence to incorrect solutions, and ultimately divergence. This happens in a regime where significant magnetic correlations develop, which are not contained in the DMFT starting point.
The convergence properties are summarized in Fig.~\ref{fig:phasediag}.
Our comparisons to numerically exact results also show that the self-consistent ladder dual fermion approximation typically yields reasonable -- and often even highly accurate -- results.
Last but not least, our DDMC data for the momentum-resolved self-energy establish a reference for benchmarking other approximate schemes for the Hubbard model.

In the future, the application of DiagMC sampling to other diagrammatic extensions of DMFT should allow for a direct comparison of the different approaches' convergence properties.
The range of parameters where the DiagMC@DF technique produces accurate converged results may be improved by considering more elaborate impurities, such as clusters, or a self-consistency scheme.

\begin{acknowledgments}
We gratefully acknowledge Evgeni Burovski for providing the Diagrammatic Determinant Monte Carlo code~\cite{DDMC_njp2006} used for the benchmark calculations. We thank E. Gull, A. A. Katanin, A. I. Lichtenstein, A.-M. Tremblay, and M. Troyer for useful discussions. JG is supported by the Swiss National Science Foundation and performed part of this work at the Aspen Center for Physics, which is supported by National Science Foundation grant PHY-1066293. EK acknowledges support of the Simons Foundation (Many Electron Collaboration program).
The DiagMC and DDMC calculations were run on the Mammouth cluster of Universit\'e de Sherbrooke, provided by the Canadian Foundation for Innovation, the Ministère de l'Éducation des Loisirs et du Sport (Québec), Calcul Québec, and Compute Canada.
The simulations and data evaluation made use of a modified version of the open source CTHYB solver~\cite{Hafermann2013} and the ALPS libraries.~\cite{alps20,alps13}
\end{acknowledgments}

\appendix

\section{Symmetrized vs.\ unsymmetrized diagrams} \label{sec:symmetrization}

\begin{figure}[tp!]
    \includegraphics[left]{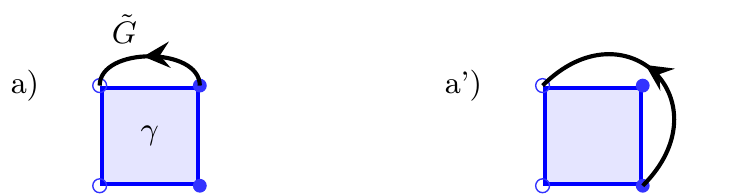}
    \includegraphics[left]{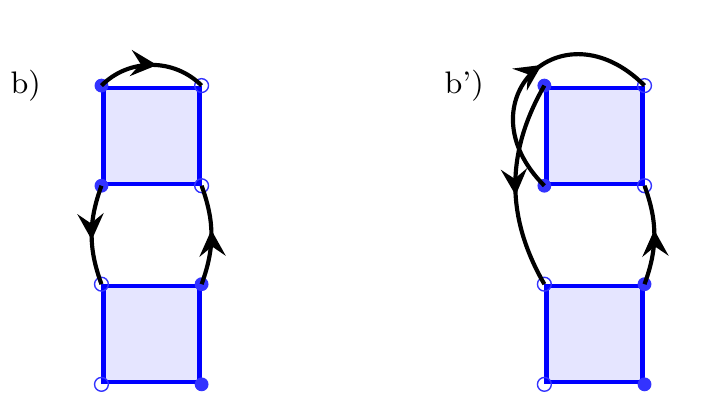}
    \includegraphics[left]{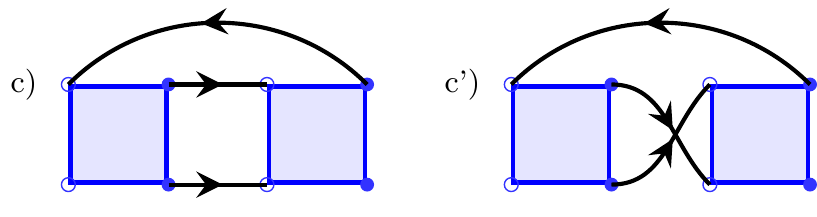}
    \caption{Diagram topologies of first- and second-order contributions to the dual self-energy. In- and outgoing corners of interaction vertices are marked with open and filled circles, respectively. In the symmetrized theory (c.f.\ Appendix~\ref{sec:symmetrization}) primed and non-primed diagrams are considered equivalent and only one variant is treated explicitly, whereas the unsymmetrized theory sums all shown diagrams.}
    \label{fig:dfdiags}
\end{figure}

Dual-fermion perturbation theory is usually described in a symmetrized formalism similar to Hugenholtz diagrams,~\cite{Hugenholtz1957,negele1988quantum} whereas our algorithm does not make use of symmetrization. It therefore sums more diagram topologies explicitly, but does not need to keep track of combinatorial prefactors associated with symmetrization. A consequence of this difference is that the value of the interaction vertex in our algorithm differs by a constant factor from the one used in other (symmetrized) DF calculations, as detailed in the following.

In the symmetrized formalism, all possibilities of connecting two interaction vertices A and B with a propagator going from A to B are considered equivalent (in other words, all corners of a vertex with an incoming line are equivalent and correspondingly for outgoing ones, c.f.\ Fig.~\ref{fig:dfdiags}). Such an approach is natural because the impurity vertex by definition is fully antisymmetric and a smaller number of distinct diagrams needs to be evaluated explicitly. For example, the single first-order diagram of the symmetrized theory accounts for both the Hartree and the Fock diagram, and the particle-hole ladder approximation does not only contain the usual ladder diagrams but also RPA-like bubble series and mixtures of these topologies.

The downside of the symmetrized theory is that each distinct topology comes with a combinatorial prefactor that needs to be determined by counting the number of equivalent lines and vertices. When arbitrary diagrams are generated by stochastic updates, this is not practical because the cost for identifying a graph's topology grows quickly with the size of the graph, in contrast to the constant cost of a local update. For DiagMC@DF sampling we therefore employ the unsymmetrized theory where the different corners of a vertex are not considered  equivalent, but the ``upper'' corners are paired with each other and correspondingly the ``lower'' ones, like in an unsymmetrized theory with some two-particle interaction $U(q)$. Then, e.g., connecting the two upper corners of a single vertex results in a Hartree diagram with a closed fermion loop, whereas the Fock diagram is a different topology obtained from connecting an upper with a lower corner.
In order to avoid double counting, the interaction of the unsymmetrized theory must differ from the reducible impurity vertex by a factor 1/2, such that the interaction of the usual DF perturbation theory is recovered by symmetrization:
\begin{align}
\gamma^{\text{symm}}(1,2; 3,4) 
&= \gamma^{\text{unsymm}}(1,2; 3,4) - \gamma^{\text{unsymm}}(1,2; 4,3) \nonumber\\&
= \frac{1}{2} \gamma^{\text{imp}}(1,2; 3,4) - \frac{1}{2} \gamma^{\text{imp}}(1,2; 4,3) \nonumber\\&
= \gamma^{\text{imp}}(1,2; 3,4).
\end{align}
Here we use the shorthand index notation $1=(x_1, \tau_1, \sigma_1)$ \etc

Furthermore, since the DiagMC updates described in Sec.~\ref{sec:updates} effectively distinguish between the upper and lower corners of the interaction vertex, whereas no such distinction is made in the diagrammatic theory (neither symmetrized nor unsymmetrized), an additional factor of 1/2 is required to avoid double counting. \footnote{
    The distinction arises from the fact that in the AIV update the upper incoming corner is always connected to the line adjacent to the worm $\mas$, and the worms $\ira$ and $\mas$ are not interchangeable. An alternative but equivalent approach is to regard $\ira$ and $\mas$ as interchangeable. Then the two worm positions are equivalent and the question of where AIV inserts the upper and lower corners is insubstantial. No additional factor of 1/2 is associated with $\gamma$, but the proposal probability of the CWO update is doubled because the excess 4-momenta $\delta$ and $-\delta$ are equivalent, so effectively the weight of the worm sector is halved. Since the insertion of an interaction is always accompanied by a transition from the worm to the physical sector, both approaches yield identical results for all physical quantities.
}
The effective interaction used in our algorithm is therefore $\gamma = \frac{1}{4} \gamma^{\text{imp}}$.

\bibliography{refs}{}

\end{document}